\documentclass[12pt]{article}
\usepackage{amsmath}
\usepackage{graphicx}
\usepackage{enumerate}
\usepackage{natbib}
\usepackage{url} 
\newtheorem{rmk}{\bf Remark}
\usepackage[caption=false]{subfig}

\newtheorem{theorem}{Theorem}
\usepackage{multirow}
\usepackage{booktabs}
 \usepackage{array} 
 \usepackage{algorithmicx}%
 \usepackage{algpseudocode}%
 \usepackage{algorithm}%
 \usepackage{amsfonts}
\usepackage{url}
\usepackage{hyperref}
\usepackage{xcolor}
\newtheorem{corr}{\bf Corollary}
\newcommand{\blind}{0}

\newtheorem{proof}{Proof}

\begin{document}




\if1\blind
{
  \title{
  \author{B Banerjee\thanks{
    The authors gratefully acknowledge \textit{please remember to list all relevant funding sources in the unblinded version}}\hspace{.2cm}\\
    Department of YYY, University of XXX\\
    and \\
    Author 2 \\
    Department of ZZZ, University of WWW}
  \maketitle
} \fi

\if0\blind
{
 
\title{
 \begin{center}
    {\large\bf 
An Efficient
Sampling from Circular Distributions and\\its Extension to Toroidal Distributions}
\end{center}}
  \author{{Surojit Biswas}\hspace{.2cm}\\ e-mail: 	surojit23@iitkgp.ac.in\\
    Department of Mathematics, Indian Institute of Technology \\
    Kharagpur, Kharagpur, 721302, West Bengal, India\\
    and \\
    Buddhananda Banerjee \\ e-mail: bbanerjee@maths.iitkgp.ac.in\\
  Department of Mathematics, Indian Institute of Technology \\
    Kharagpur, Kharagpur, 721302, West Bengal, India\\
    and}
  \maketitle

  \medskip
} \fi

\begin{abstract}
Sampling from circular distributions is a fundamental task in directional statistics. A key challenge in acceptance–rejection methods lies in selecting an efficient envelope density, as poor choices can lead to low acceptance rates and increased computational cost, especially in large-scale simulations. To address this, we propose a new sampling framework that utilizes the idea of upper Riemann sums to construct a piecewise envelope. This method ensures validity for any Riemann-integrable target density on a bounded interval. This method exhibits enhanced efficacy relative to the present sampling method for the von Mises distribution.
Additionally, we introduce a flexible family of distributions defined on the surface of a curved torus, using its area element. The proposed sampling method is then employed to generate samples from the toroidal model. We explore the maximum entropy characterization and other theoretical properties of one of the marginal distributions arising from this construction for the von Mises distribution.
To illustrate the practical utility of our framework, we apply the model to a real dataset on wind direction.

\end{abstract}

\noindent%

{\it Keywords:}  Acceptance-Rejection sampling, Upper Riemann sum, Circular distribution, Curved torus, Area element, Toroidal-distributions.
\vfill

\newpage

\section{Introduction}
Sampling from circular distributions, such as the von Mises, cardioid, and Kato–Jones distributions, is a vital challenge in directional statistics, with the acceptance-rejection method being one of the most commonly utilized techniques. A significant problem in this methodology is the selection of a suitable envelope (or proposal) distribution, as the acceptance percentage and thus computing efficiency greatly rely on this decision.  In extensive simulations, an inappropriate envelope might cause high rejection rates, leading to a considerable computational burden.

This study presents a new and comprehensive methodology for developing efficient envelope distributions utilizing the concept of upper Riemann sums.  This method significantly decreases the number of rejected samples while maintaining the authenticity of the sampling process.  The suggested method is applicable to any target probability density function that is Riemann integrable across a bounded interval. This encompasses a wide range of circular distributions of interest.  Utilizing the step-function framework of the upper Riemann sum, we develop a dominating function that can be normalized to create a legitimate proposal density.  This offers an effective and theoretically justified method for enhancing the acceptance rate in rejection sampling. The proposed sampling method, when applied to the von Mises distribution, showcases a notably improved acceptance rate and reduced runtime compared to the current method by \cite{best1979efficient}.


This paper also focuses on the 2-dimensional curved torus $(\mathbb{S}^1 \times \mathbb{S}^1)$, a Riemannian manifold embedded in \( \mathbb{R}^3 \). Circular and spherical data are also modeled as Riemannian manifolds and naturally incorporate the geometry of their respective surfaces in statistical analysis \cite[see][]{mardia2000directional}. On the contrary to that, the flat torus, $(0, 2\pi] \times (0, 2\pi]$, does not offer a similar geometric structure. There are several studies have been done on the distributions over the flat torus e.g., by \cite{mardia1975statistics, rivest1988distribution, singh2002probabilistic, mardia2000directional, ameijeiras2022sine}. We prefer to represent bivariate angular data on the curved torus because the flat and curved tori are not homeomorphic due to their topological properties. This emphasizes the importance of incorporating the appropriate geometric framework when analyzing toroidal data. The curved torus enjoys smooth identifiability of the edges, leading to a compact and connected support of a distribution on it.

The von Mises distribution on the circle and the Fisher distribution on the sphere are intrinsic distributions that heavily rely on their respective spaces' length element and area element. Both distributions are characterized as maximum entropy distributions within their respective spaces, subject to certain constraints. 
In a similar vein, inspired by the research of \cite{diaconis2013sampling}, we propose a distribution on a curved torus that also arises from its intrinsic geometry.  Furthermore, it is a natural extension of the von Mises distribution evolved with the area element of the curved torus, providing one of the marginal distributions on it. Whereas the other marginal can be suitably chosen as the von Mises distribution or the uniform circular distribution.
Analogous to the von Mises distribution on the circle and the Fisher distribution on the sphere, we demonstrate that under specific constraints, the proposed distribution emerges as the \textit{maximum entropy} distribution on the surface of the curved torus.

Now, to get the random sample from the proposed toroidal distributions, we generate it individually from each of the marginals and assemble them component-wise. To generate random samples from the distribution on the horizontal circle (associated with the larger radius $R$) and that from the distribution on the vertical circle (associated with the smaller radius $r$), we implement the proposed modified acceptance-rejection sampling technique. This proposed sampling scheme will accelerate the empirical inferential studies for a large-scale simulation.\\

Our investigation is structured into several sections.  In Section \ref{ch:HAR}, we present an effective simulation methodology for generating random samples from many circular distributions. 
 In Section \ref{general_dist_torus}, we present the formulation of a general distribution on the surface of a curved torus.  The sampling methodologies derived from the marginal distributions on the horizontal and vertical circles are illustrated in Section \ref{sampling_on_torus}.

Section \ref{voncos_properties} focuses on the study of the modified von Mises distribution as the marginal on the vertical circle of its toroidal counterpart, starting with some special cases.  Section \ref{max_entropy} examines the natural constraints of the corresponding maximum entropy distribution on the torus.  Section \ref{trig_moments} presents the derivation of closed-form expressions for trigonometric moments.  Sections \ref{condition_symm} to \ref{div_from_cardiod} examine essential distributional properties, such as symmetry, (bi)modality, and divergence from the cardioid distribution.  Section \ref{MLE} outlines the maximum likelihood estimation procedure, incorporating observed and expected information matrices, and emphasizes the necessity for general numerical optimization techniques.
 Section \ref{symm_case} is devoted to the examination of the two-parameter family of symmetric and unimodal distributions with a location parameter $\mu=0$. 
 The concluding Section-\ref{conclusion} follows the necessary proofs in Appendix \ref{appendix} and is preceded by a comprehensive analysis of real data, especially the wind direction data presented in Section-\ref{data_analysis}. In the following we provide a description of the data.

\textbf{Wind Direction Data:}
    The direction of wind plays a pivotal role in shaping meteorological phenomena and climatic dynamics, impacting a myriad of sectors ranging from agriculture to urban planning and air quality management. Recognizing the intricacies of wind direction variability is paramount for applications such as renewable energy production, where wind patterns dictate the feasibility and efficiency of wind power generation. In this research paper, we delve into the modeling of wind direction variability specifically at Kolkata (Latitude 22.57, Longitude 88.36), the capital city of West Bengal, India. Spanning over four decades, from 1982 to 2023, our study focuses on the month of August, meticulously examining trends and fluctuations in wind direction to elucidate its implications on local weather patterns and climatic conditions. 


\section{  An efficient sampling from circular distributions}
\label{ch:HAR}
In this section, we present a modified acceptance–rejection sampling methodology designed to improve efficiency when sampling from angular distributions. The central idea is to construct a piecewise constant envelope function based on the upper Riemann sum of the target density. The method applies broadly to any Riemann-integrable density on a bounded interval and is especially well-suited for circular distributions such as the von Mises and Kato–Jones families. By normalizing the upper Riemann sum, we obtain a valid proposal distribution that enables efficient and theoretically sound sampling. The following subsections provide the formal construction and algorithmic details.

\subsection{Proposed  sampling method}
\label{sampling_method}
Let $U$ follow a uniform distribution on $[0,1]$.
Assume that $f(x)$ and $p(x)$ are the target and proposed probability density functions, respectively, with the common finite support $[a,b]$. Consider a  partitions of the interval $[a,b]$ as: $a=a_0<a_1<a_2<\cdots<a_{i-1}<a_i<\cdots<a_{k-1}<a_k=b$.
Now, for $i=1,\cdots, k$, let 
$ A_{i} = [a_{i-1},a_i],$ $ B = a_{i}-a_{i-1}=\frac{(b-a)}{k}, $ and $P(A_i) = \displaystyle \int_{A_i}f(y) dy $. Now, to obtain samples from the target density, we first choose one of the intervals corresponding to its probability and then use rejection sampling over the conditional distribution of that interval with a uniform envelope associated with the maximum height of the function on that interval. Mathematically, we demonstrate the same.
 Let, for the $i^{th}$ cell $H_i=\max\limits_{x\in A_i}
 f(x)$, $Y_i=a_{i-1}+BU$. It is easy to note that $Y_i \sim U[a_{i-1}, a_i]$. So, the proposed density for the entire support is 
 \begin{equation}
     p(y)= \sum_{i=1}^{k} \dfrac{H_{i} \textbf{I}_{A_i}(y)}{\displaystyle B\sum_{i=1}^{k} H_i}
     \label{proposed density}
 \end{equation}
 satisfying the condition that $p(y \mid y\in A_i) =\frac{1}{B}$. 
Since all intervals $A_i = [a_{i-1}, a_i]$ have equal width $B = (b-a)/k$, and the proposal density within each interval is uniform, the conditional density $p(y \mid y \in A_i)$ is constant and equal to $1/B$ for all $y \in A_i$.
 Once a subinterval $A_i$ is selected and a candidate $Y_i$ is generated, we accept the sample $Y_i$ with probability
$\frac{f(Y_i)}{H_i}.$
This ensures that the piecewise-constant proposal density function, $ p(y)$ defined in Equation~\eqref{proposed density} can be used in the acceptance–rejection procedure.
Now, we choose the number, $M_i=\frac{BH_i}{P(A_i)}$ for the $i^{th}$ cell such that $M_i\geq \max\limits_{y\in A_i}\frac{f(y\mid y\in A_i)}{p(y \mid y\in A_i)}.$
As a consequence,

\begin{eqnarray}
      \displaystyle  \int_{a}^{x} f(t)\,\,dt &=& \sum_{i=1}^{k} P(A_{i}) \left[ P \left(Y_i\leq x \,| \,M_i\geq\max\limits_{y\in A_i}\frac{f(y\mid y\in A_i )}{p(y \mid y\in A_i )}\right)\right],\nonumber\\
       &=& \sum_{i=1}^{k} P(A_{i}) \left[ P(Y_i\leq x \,| \,Y_i  \hspace{.2cm}\text{accepted})\right] \nonumber
\end{eqnarray} see \ref{appendix acceptence} for detailed calculation.
This method can be implemented in 
 the following Algorithm-\ref{alg:algo HAR}, which is the pseudo-code of the proposed sampling method.\\


%

\begin{algorithm}
\caption{ The proposed sampling algorithm.}\label{alg:algo HAR}
\begin{algorithmic}[1]
\Require Target probability density function $f(x)$ with support $[a,b]$.
\Ensure Samples from the probability density function $f(x)$ with support $[a,b]$.

\State$n \Leftarrow$ Number of random samples to be generated\;
\State$np \Leftarrow$ Number of partitions with equal length\;
\State$pt\Leftarrow$  A sequence in $[a,b]$ with length $(np+1)$ \;
\State$H\Leftarrow f(pt)$ \Comment{ Heights of the probability density function at pt}
\State$H_m\Leftarrow$ A sequence of the maximum heights of each partition \;
\State$bl\Leftarrow \frac{b-a}{np}$ \Comment{ Bin length } 
\State$p_m \Leftarrow \displaystyle \frac{H_m}{\sum H_m}$\Comment{ Probability vector}
  \State$y \Leftarrow \mbox{~Initialize an empty vector of size n} $\;
\State $i \Leftarrow 0$
\While{$i \leq n$}
\State$u\Leftarrow$ Draw a number random  from $U[0,1]$ \;
\State $ml \Leftarrow $ Draw a random sample from multinomial with probability vector $p_m$ \;
\State $x \Leftarrow pt[ml]+u*bl$\;
\State $px\Leftarrow \frac{f(x)}{H_m[ml]}$\;
\State$rp \Leftarrow$ A random number from $Bernoulli(px)$\;
  \If{  $rp=1$ }
            \State      $i\Leftarrow i+1$\;
             \State $y[i]\Leftarrow$  $x$\; \Comment{  Accepted sample}
        \EndIf
\EndWhile
\end{algorithmic}
\end{algorithm}



\textbf{von Mises:} Here, we demonstrate the superiority of the proposed sampling scheme over the existing one for the well-known von Mises distribution having 
\label{vm_simulation}
The probability distribution function 
\begin{equation}
    f_{vm}(\theta)=\frac{e^{\kappa\cos(\theta-\mu)}}{2\pi I_{0}(\kappa)},
\label{von mises}
\end{equation}
where $0\leq \theta<2\pi$, $0\leq \mu<2\pi$, $\kappa>0$, and $ I_{0}(\kappa)$ is the modified Bessel function with order zero evaluated at $\kappa,$ defines the von Mises distribution. It is a widely applicable probability distribution for modeling angular or circular data. From Equation (\ref{von mises}), it is evident that the probability distribution with concentration parameter $\kappa$ is a continuous and symmetric distribution with respect to the mean direction $\mu$. 
Sampling from the von Mises distribution is challenging because its cumulative distribution function (CDF) has no closed form, as discussed in \cite{mardia2000directional}. Therefore, this distribution cannot use conventional sampling methods such as inverse transform sampling.\

The sampling procedure proposed by \cite{best1979efficient} for the von Mises distribution is a well-known method that uses the conventional rejection sampling technique with the wrapped Cauchy distribution as an envelope, which we refer to as vMBFR sampling. However, depending on the parameter, the vMBFR sampling method has a high rejection rate. On the contrary, the proposed sampling method has a much lower rejection rate. \

In this study, we conducted a simulation with a sample size $n=50000$, partition $k=250$ to compare the acceptance percentage of sample size between the proposed and vMBFR algorithms.  Table-\ref{table:vm HAR1} and Table-\ref{table:vm HAR2}  present the acceptance percentage for different values of the concentration parameter $\kappa$ with a mean direction parameter $\mu=0$. The proposed sampling method outperforms the existing vMBFR sampling method in terms of the acceptance percentage of sample size. 
The run-time of the proposed and vMBFR algorithms has been compared per $10^6$ sample. 
The mean run-time (with standard deviation) for $10^3$ iterations has been reported in  Table-\ref{table:run-time}. A system with a processor$Intel \circledR ~ Xeon(R)~ CPU ~E5-2630~ v3~ @ ~2.40GHz \times 16$, RAM $64.0$ GB, graphics$NVS~ 315/PCIe/SSE2$, Ubuntu 22.04.2 LTS, $64$-bit OS has been used for the above simulation. A comparison of run-time between the proposed $100, 250, 500$ partitions and vMBFR algorithms for generating data from von Mises distribution with different parameters $\kappa$ has been reported. It is observed that the proposed algorithms take substantially less time compared to that of the vMBFR. \

Figure-\ref{vm_kj}(a) displays the histogram of the sampled points from the von Mises distribution using the proposed sampling technique.
\begin{table}[h!]
\centering
\begin{tabular}{|l|c|c|c|c|c|c|c|c|c|c|}
\hline
$\kappa$ & $0.1$ & $0.2$ & $0.3$ & $0.4$ & $0.5$ & $0.6$ & $0.7$ & $0.8$ & $0.9$ & $1$\\
\hline
\hline
\textbf{Proposed} & $99.96$ & $99.92$ & $99.87$ & $99.85$ & $99.81$ & $99.77$ & $99.72$ & $99.71$ & $99.67$ & $99.65$\\

\hline
\textbf{vMBFR } & $99.76$ & $99.06$  & $97.90$ & $96.67$ & $95.04$   & $93.23$  & $91.88$  & $89.88$ & $88.12$ & $86.94 $\\
\hline
\end{tabular}
\vspace{.051cm}
\caption{Acceptance percentage comparison  for von Mises distribution with $ 0.1\leq \kappa \leq 1$ for von Mises distribution.}
\label{table:vm HAR1}
\end{table}

\begin{table}[h!]
\centering
\scalebox{1}{
\begin{tabular}{|l|c|c|c|c|c|c|c|c|c|c|}
\hline
$\kappa$ & $2$ & $3$ & $4$ & $5$ & $10$ & $20$ &$40$ & $60$  &$80$ &$100$ \\
\hline
\hline
\textbf{Proposed } & $99.48$ & $99.21$ & $99.02$ & $98.91$ & $98.462$ & $97.76$ &$96.96$ & $96.31$ &$96.76$ & $95.15$ \\

\hline
\textbf{vMBFR } & $76.95$ & $72.37$  & $69.96$ & $69.46$ & $67.46$   & $66.64$  & $66.43$  & $65.96$ &$65.94$ &$65.69$\\
\hline
\end{tabular}}
\vspace{.1cm}
\caption{Acceptance percentage comparison for von Mises distribution with  $ 2\leq \kappa \leq 100$ .}
\label{table:vm HAR2}
\end{table}

\begin{table}[!h]
\scalebox{0.95}{
\renewcommand{\arraystretch}{1.2}
\begin{tabular}{|l||l l|l l|l l|}
\hline
   &\multicolumn{2}{|c|}{$k=100$ partitions }&\multicolumn{2}{|c|}{$k=250$ partitions }&\multicolumn{2}{|c|}{$k=500$ partitions }\\

$\boldsymbol{\kappa}$ &\textbf{Proposed}&\textbf{vMBFR} &\textbf{Proposed}&\textbf{vMBFR}&\textbf{Proposed}&\textbf{vMBFR}\\
\hline\hline
0.1& 3.58(0.12) &6.23(0.04)&3.46(0.13) &6.23(0.04)&3.51(0.15)&6.29(0.13)\\

0.5&3.60(0.18) &6.29(0.13)&3.53(0.13)&6.29(0.13)&3.51(0.09) &6.29(0.13)\\

1&3.57(0.16)&6.29(0.09) &3.51(0.12)&6.29(0.09) &3.47(0.14)&6.29(0.09) \\

5&3.64(0.19)&6.27(0.04)&3.60(0.07)&6.27(0.04)&3.45(0.13)&6.27(0.04)\\

10&3.75(0.15)&6.28(0.04) &3.64(0.11) &6.28(0.04)&3.56(0.17) &6.28(0.04)\\

20&3.64(0.08)&6.25(0.13)& 3.44(0.07)&6.25(0.13)&3.41(0.09)&6.25(0.13)\\

50&4.12(0.15)&6.31(0.15) &3.71(0.19)&6.31(0.15) &3.52(0.13)&6.31(0.15) \\

100&4.43(0.15)&6.30(0.07)& 3.96(0.17)&6.30(0.07)&3.56(0.11)&6.30(0.07)\\
\hline
\end{tabular}}													
\vspace{0.15cm}
\caption{Run-time comparison between the proposed with $100, 250, 500$ partitions, and vMBFR algorithms for generating data from von Mises distribution with different parameters $\kappa.$ Average run-time (with standard deviation) in seconds is reported for a sample size of $10^6$ with $10^3$ iterations each.}
\label{table:run-time}					 						
\end{table}

\textbf{Kato-Jones:} Another well-known new circular distribution is the Kato and Jones distribution. It is a four-parameter family of circular distributions that was first introduced by \cite{kato2010family} where they have used the M\"{o}bius transformation. The probability density function of the distribution is given by
\begin{eqnarray}
      f_{kj}(\theta)= \frac{(1-\rho^2)(2\pi I_{0}(k))^{-1}}{1+\rho^2-2\rho \cos{(\theta-\gamma)}} \exp\left[ \dfrac{\kappa \{ \xi \cos{(\theta-\eta)}-2\rho \cos \nu_1 \}}{1+\rho^2-2\rho \cos{(\theta-\gamma)}}   \right],
    \label{kato jons}
\end{eqnarray}

where $0\leq \mu,\nu_1<2\pi$, and $0\leq \rho <1$, $\kappa>0$, and 
$\gamma=\mu+\nu_1$, $\xi=\sqrt{\rho^4+2\rho^2\cos{(2\nu_1)}+1}$, $\eta=\mu+\arg(\rho^2\cos{(2\nu_1)}+1+i\rho^2\sin{(2\nu_1)})$.

Random variate generation from this distribution is derived directly from the density construction detailed in Equation~(\ref{kato jons}), as described by \citet{kato2010family}.  They generated samples by applying an M\"{o}bius transformation to random variates from the von Mises distribution as outlined by \citet{best1979efficient}, previously described in this section.  Thus, the acceptance rate of this method aligns with that of the conventional von Mises sampling technique; however, it entails an extra computational expense owing to the added transformation step.  The proposed sampling algorithm demonstrates a significantly higher acceptance rate compared to the method presented by \citet{kato2010family}, alongside a decrease in run-time.  The acceptance percentages achieved by the proposed method for $n=50000$ and partition $k=250$ at fixed values of $\rho$ and $\kappa$ are presented in Tables \ref{table:kj rho fix HAR2} and \ref{table:kj kp fix HAR2}, respectively.  To maintain brevity, the associated run-time results are excluded.  Figure-\ref{vm_kj}(b) presents the histogram of random samples derived from the Kato–Jones distribution, defined by the density in Equation (\ref{kato jons}), with parameter values $\mu = \frac{\pi}{3}, \nu_1 = \frac{\pi}{2}$, $\rho = 0.3$, and $\kappa = 1$. Since the Kato-Jones distribution is asymmetric, for the implementation, the maximum height $H_i=\max\limits_{x\in A_i}
 f_{kj}(x)$
  over each subinterval $A_i=[a_{i-1},a_i]$ is replaced by $H_i^*= f_{kj}\left(\frac{a_{i-1}+a_i}{2} \right)$, which provides a computationally efficient and reasonably closed envelope for the sampling. Such modification can be implemented for any other asymmetric distribution.

\begin{table}[h!]
\centering
\scalebox{0.9}{
\begin{tabular}{|l|c|c|c|c|c|c|c|c|c|c|}
\hline
$\kappa$ & 1 &2& 3& 4& 5 &6& 7 &8 &9 &10 \\
\hline
\hline
\textbf{Proposed } &98.742& 98.078 &97.502& 97.084& 96.756 &96.448 &96.298& 96.098& 95.604 & 94.864\\
\hline
\end{tabular}}
\vspace{.1cm}
\caption{Acceptance percentage for the density  given in Equation (\ref{kato jons}) with  $\mu = \frac{\pi}{3}, \nu_1 = \frac{\pi}{2}$, $\rho=0.5$  (fixed) and varying $\kappa$.}
\label{table:kj rho fix HAR2}
\end{table}

\begin{table}[h!]
\centering
\scalebox{1}{
\begin{tabular}{|l|c|c|c|c|c|c|c|c|c|c|}
\hline
$\rho$ & 0.1 &0.2& 0.3& 0.4& 0.5 &0.6& 0.7 &0.8 &0.9  \\
\hline
\hline
\textbf{Proposed } & 99.496 &99.414 &99.250 &99.072& 98.710& 98.352& 97.598 &96.438& 92.424 \\
\hline
\end{tabular}}
\vspace{.1cm}
\caption{Acceptance percentage for the density  given in Eqiuation (\ref{kato jons}) with  $\mu = \frac{\pi}{3}, \nu_1 = \frac{\pi}{2}$, $\kappa=1$  (fixed) and varying $\rho$.}
\label{table:kj kp fix HAR2}
\end{table}

\begin{figure}[h!]
\centering
\subfloat[]{%
{\includegraphics[trim=0 60 60 60, clip,width=0.5\textwidth,height=0.5\textwidth]{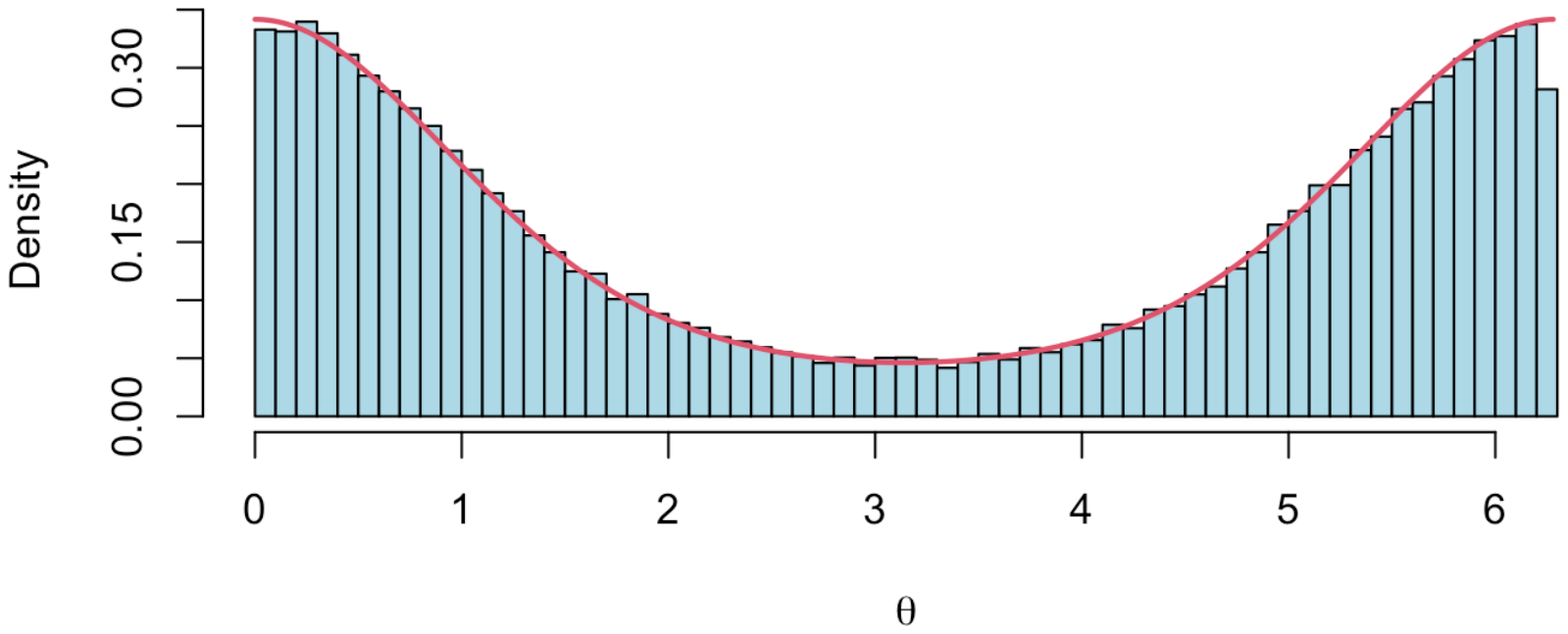}}}
\subfloat[]{%
{\includegraphics[trim=0 60 60 60, width=0.5\textwidth,height=0.5\textwidth]{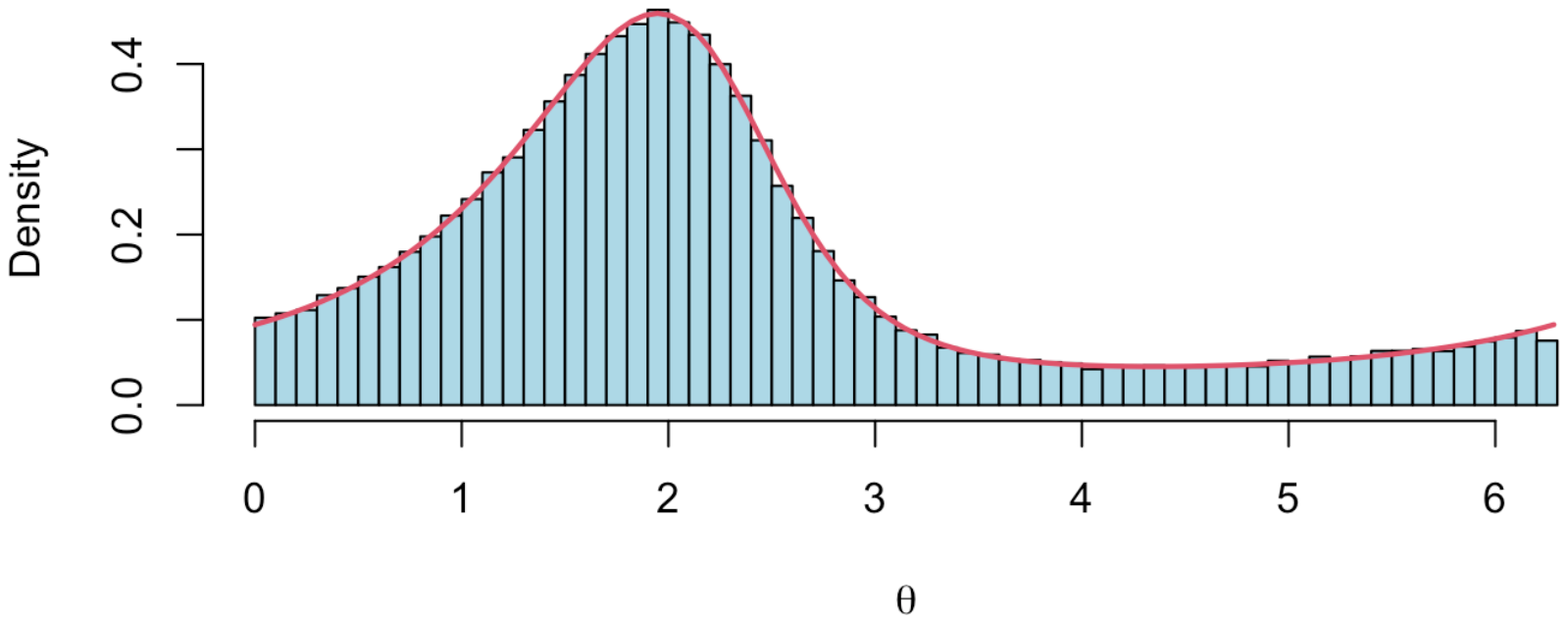}}}
\caption{ (a) Histogram of the random samples from von Mises distribution with $\mu=0, \kappa=1.$ (b) Histogram of the random samples from the Kato-Jones distribution with $\mu = \frac{\pi}{3}, \nu_1 = \frac{\pi}{2}$, $\rho = 0.3$, and $\kappa = 1$.} 
\label{vm_kj}
\end{figure}

\section{Construction of general distribution on the surface of the curved torus} \label{general_dist_torus}
A torus is a geometric object representing two angular variables with respective radii. 
Only the angular part of it can be represented with a flat torus $[0,2\pi)\times [0,2\pi)$, but when the radii are involved, it is represented as the curved torus, see Equation (\ref{torus equation}). The curved torus is not homeomorphic to the flat torus because their topological properties differ. To analyze the data represented on the surfaces of a curved torus, it is essential to have a proper notion of probability distributions it and statistical methodologies for inference. The existing statistical techniques from the literature applicable to the flat torus do not apply to the analysis of data on a curved torus because it does not take into account the topology and geometry of the surface. \\

Here, we focus on the  $2$-dimensional curved torus, a Riemannian manifold embedded in the $\mathbb{R}^{3}.$ In this report, we will use the term ``curved torus" for $2$-dimensional curved torus. The parameter space for the curved torus is $\mathcal{S}=\{ (\phi, \theta): 0\leq \phi, \theta<2\pi  \}$, and that can be represented in parametric equations as
 \begin{equation}
 \begin{aligned}
      x(\phi,\theta) &= (R+r\cos{\theta})\cos{\phi}\\
    y(\phi,\theta) &= (R+r\cos{\theta})\sin{\phi} \\
    z(\phi,\theta) &= r\sin{\theta},
 \end{aligned}
 \label{torus equation}
 \end{equation}
   where  $R,r$ are radii of the horizontal and vertical circles, respectively. 

Although several distributions on the flat torus have been introduced and studied by many researchers, only the distribution on the surface of the curved torus has been investigated by \cite{diaconis2013sampling}. They have specially studied uniform distribution on the torus, as given in Equation (\ref{torus equation}) as an example of a manifold.
In this section, we introduced a generalization for some popular distributions on the surface of a curved torus as an extension of the circular distribution in higher dimensions.

We can determine the area element of the torus as $dA=r(R+r\cos{\theta})~ d\phi d\theta$ (see Appendix-\ref{geometry_torus}). Hence, it is immediate that the area of the torus is given by
\begin{equation}
    A = \int_{0}^{2\pi} \int_{0}^{2\pi}  r(R+r\cos{\theta})\,\,d\phi\,d\theta=4\pi^2rR
\label{area torus}
\end{equation}
Now, let us consider $\frac{r}{R}=\nu$, and a joint probability density function, $h(\phi,\,\theta)$ of $\phi$ and $\theta$, and from the Equation (\ref{area torus}) obtain the identity

\begin{eqnarray}
 4\pi^2rR &=& \int_{0}^{2\pi} \int_{0}^{2\pi} h(\phi,\,\theta) (2\pi r)(2\pi R)\,d\phi\,d\theta \nonumber\\
     &=& \frac{1}{C}\int_{0}^{2\pi} \int_{0}^{2\pi} h(\phi,\,\theta) \left(1+\nu\cos\theta \right)
     (2\pi r)(2\pi R)\,d\phi\,d\theta,\nonumber
\end{eqnarray}
where $C$ is a normalizing constant, implying

\begin{eqnarray}
      1&=& \frac{1}{C}\int_{0}^{2\pi} \int_{0}^{2\pi} h(\phi,\,\theta) \left(1+\nu\cos\theta \right)
     \,d\phi\,d\theta\nonumber\\
      &=&   \frac{1}{C}  \int_{0}^{2\pi} \int_{0}^{2\pi} h_{1}(\phi \,|\, \theta)\, \left[\,h_{2}(\theta)\left(1+\nu\cos\theta \right)\right]
     \,d\phi\,d\theta.
   \end{eqnarray}
Here, in general, we can consider  sampling from the joint probability density function

\begin{equation}
    h^{*}(\phi,\theta)=\frac{1}{C}\, h_{1}(\phi \,|\, \theta)\, \left[\,h_{2}(\theta)\left(1+\nu\cos\theta \right)\right], \mbox{~where~~} 0\leq \phi, \theta<2\pi
      \label{torus area dist}
\end{equation}
as a sampling scheme from the surface of a curved torus where the joint density on the parameter space or in the flat torus is pre-specified as $h(\phi, \theta).$ In particular, when $\phi$ and $\theta$ are independently distributed then $ h_{1}(\phi \,|\, \theta)\propto h_{1}(\phi)$. Hence, the Equation (\ref{torus area dist}) will reduce to
\begin{equation}
    h^{*}(\phi,\theta)\propto h_{1}(\phi)\, \left[h_{2}(\theta)\left(1+\nu\cos\theta \right)\right], \mbox{~where~~} 0\leq \phi, \theta<2\pi.
    \label{torus independent dis}
\end{equation}
The above decomposition facilitates us to generate random samples individually from $h_{1}(\phi)$, and $h_{2}(\theta)\left(1+\nu\cos\theta \right)$ up to the normalizing constant for the horizontal circle and the vertical circle, respectively.  

\begin{rmk}
   It is important to note that the marginal density associated with the 
    vertical circle naturally possesses the multiplicative factor $\left(1+\nu\cos\theta \right)$ due to the area element of the torus with radius ratio $\nu=\frac{r}{R} \in(0,1].$ It is not an externally perturbed circular distribution. Hence, the parameter(s) of the density $h_2(\theta)$ do not influence the multiplicative factor $\left(1+\nu\cos\theta \right)$. For the spherical distribution, a similar multiplicative factor ($\sin \theta$) is also present, which is free from the parameters of the prevailing distribution \cite[see][ch. 9]{mardia2000directional}.
\end{rmk}

\begin{rmk}
   Note that the marginal distribution, $h_{1}(\phi)$ of horizontal angle $\phi$ can be any circular distribution. We can use the proposed sampling method to generate random samples for $h_{1}(\phi)$.
The marginal distribution, $h_{2}(\theta)$ of the vertical angle $\theta$, is a non-standard distribution on the circle which involves the area element of the curved torus. Hence, in the following section, we provide the sampling method from $h_{2}(\theta)$ as an application of the proposed sampling method in Section-\ref{sampling_method}. 
\end{rmk}

\section{ Sampling from the marginal density on vertical cross-sectional circle of the torus}
\label{sampling_on_torus}
As we know, the marginal distribution on the horizontal circle will be proportional to $h_1(\phi)$ from Equation (\ref{torus independent dis}). In contrast, the distribution of the vertical circle will be proportional (or equal) to $h^{*}_{2}(\theta)=\left[h_{2}(\theta)\left(1+\nu\cos\theta \right)\right]$   after modifying by the area element of the curved torus. The newly proposed Algorithm-\ref{alg:algo HAR} in Section-\ref{ch:HAR} will facilitate us in generating samples from both the marginal distributions.

\begin{enumerate}

\item
\textbf{von Mises Distribution:}
In this section, we proposed the extension of the von Mises distribution on the surface of the curved torus. Let $\phi $ and $\theta$ be independently von Mises distributed on the flat torus with concentration parameters $\kappa_1$, $\kappa_2$, and location parameters $\mu_1$, $\mu_2$, respectively.  So, now, using Equation (\ref{torus area dist}) we can define the following new distribution as

\begin{equation}
    h^{*}(\phi,\theta)= \frac{e^{\kappa_1\cos(\phi-\mu_1)}}{2\pi I_{0}(\kappa_1)} \left[ \frac{e^{\kappa_2\cos(\theta-\mu_2)}}{C} \left(1+\nu\cos\theta \right)\right], 
    \label{von torus}
\end{equation}
where $0\leq \phi, \theta<2\pi$, $0\leq \mu_1,\mu_2<2\pi$, $\kappa_1,\kappa_2>0$, which implies 

$$ h_{1}(\phi)=\frac{e^{\kappa\cos(\phi-\mu_1)}}{2\pi I_{0}(\kappa_1)} \mbox{~and~}
 h_{2}(\theta)=\frac{e^{\kappa_2\cos(\theta-\mu_2)}}{C} \left(1+\nu\cos\theta \right),$$
with $$C=2\pi\left[I_{0}(\kappa_2)+\nu \cos \mu_2~I_{1}(\kappa_2)\right]$$ 
is the normalizing constant (see Appendix-\ref{appendix har}). 


The probability distribution function of the vertical angle $\theta$ is 

\begin{equation}
    h_2(\theta)=\frac{e^{\kappa\cos(\theta-\mu)}\left(1+\nu\cos\theta \right)}{2\pi(I_{0}(\kappa)+\nu \cos \mu~I_{1}(\kappa))},
    \label{voncos}
\end{equation}
where, $\theta,\mu \in[0,2\pi)$, $\kappa>0$, and $\nu\in (0,1).$
   
     


\label{vertical_graph_density}
Now let $\mu=\frac{\pi}{3}$, $\kappa=1$, and $\nu=0.5$, then the below Figure-\ref{vm}(a) is the histogram of the sampled data drawn from the distribution of the vertical angle $\theta$ using the newly proposed sampling method in Section \ref{sampling_method}, and the acceptance rates for $n=50000$ and partition $k=250$ associated with this method are summarized in Table-\ref{table:voncos HAR2}.


\begin{table}[h!]
\centering
\scalebox{0.9}{
\begin{tabular}{|l|c|c|c|c|c|c|c|c|c|c|}
\hline
$\kappa$ & 1 &2& 3& 4& 5 &6& 7 &8 &9 &10 \\
\hline
\hline
\textbf{Proposed } & 99.456& 99.430& 99.498 &99.434& 99.438& 99.478& 99.440& 99.468& 98.428& 98.228\\
\hline
\end{tabular}}
\vspace{.1cm}
\caption{Acceptance percentage for the density $h_2(\theta)$ with  $\mu=\frac{\pi}{3}$, $\nu=0.5$, and varying $\kappa$.}
\label{table:voncos HAR2}
\end{table}


\item 
\textbf{Wrapped Cauchy Distribution:}
The wrapped Cauchy distribution is one of the well-known circular distributions given by the probability density function
\begin{equation}
    f_{wc}(\theta)=\frac{1}{2\pi}\dfrac{1-\rho^2}{1+\rho^2-2\rho \cos({\theta-\mu})},
    \label{wrap cauchy}
\end{equation}
where $0\leq \theta<2\pi$, $0\leq \mu<2\pi$, and $0\leq \rho <1$.

Now, we present the extension of the wrapped Cauchy distribution on the surface of the curved torus.  Let $\phi $ and $\theta$ are independently wrapped Cauchy distributed with concentration parameter, $\rho_1,\rho_2$, and location parameter $\mu_1,\mu_2$, respectively on flat torus.  Therefore, the corresponding joint density is given by

\begin{equation}
    h^{*}(\phi,\theta)=\dfrac{(2\pi)^{-1}(1-\rho_1^2)}{1+\rho_1^2-2\rho_1\cos({\phi-\mu_1})} \left[ \frac{(C~2\pi)^{-1}(1-\rho^2_2 )}{1+\rho_2^2-2\rho_2\cos({\theta-\mu_2})} \left(1+\nu\cos\theta \right)\right], 
    \label{wc torus}
\end{equation}
where $0\leq \phi, \theta<2\pi$, $0\leq \mu_1,\mu_2<2\pi$, $0\leq \rho_1,\rho_2<1$, which implies
$$ h_{1}(\phi)= f_{wc}(\phi) \mbox{~and~} h^{*}_{2}(\theta)=\frac{1}{C}f_{wc}(\theta)\left(1+\nu\cos\theta \right),$$
 with a normalizing constant
$$C=\displaystyle \int_{0}^{2\pi} \left[ f_{wc}(\theta)~\left(1+\nu\cos\theta \right) \right]~d\theta.$$

Hence, the joint probability density function
in Equation (\ref{wc torus}) is the representative of the wrapped Cauchy distribution on the surface of a curved torus. Now, let us assume $\mu_2=0$, $\rho_2=0.5$, and $\nu=0.5.$ 
Figure-\ref{vm}(b) presents the histogram of sampled values for the vertical angle $\theta$, obtained using the proposed sampling method described in Section-\ref{ch:HAR}. The acceptance rates for $n=50000$, and partition $k=250$ associated with this method are summarized in Table-\ref{table:wc  fix HAR2}.

\begin{table}[h!]
\centering
\scalebox{1}{
\begin{tabular}{|l|c|c|c|c|c|c|c|c|c|c|}
\hline
$\rho_2$ & 0.1 &0.2& 0.3& 0.4& 0.5 &0.6& 0.7 &0.8 &0.9  \\
\hline
\hline
\textbf{Proposed } &99.710& 99.584& 99.520& 99.398& 99.290& 99.084& 98.772& 98.154 &96.090 \\
\hline
\end{tabular}}
\vspace{.1cm}
\caption{Acceptance percentage for the density $\frac{1}{C}f_{wc}(\theta)\left(1+\nu\cos\theta \right)$ with  $\mu_2=0$, and varying $\rho_2$ .}
\label{table:wc  fix HAR2}
\end{table}

\item  

\textbf{Kato and Jones Distribution:}

Now, we present the Kato and Jones distribution with the density given in Equation (\ref{kato jons}) on the surface of the curved torus.  Let $\phi $ and $\theta$ are independently Kato and Jones distributed with concentration parameters, $\rho_1,\rho_2,\kappa_1, \kappa_2$, and location parameter $\mu_1,\mu_2,\nu_1,\nu_2$, respectively on flat torus.  Therefore, the corresponding joint density on the surface of the curved torus is given by
\begin{multline}
    h^{*}(\phi,\theta)= \frac{(1-\rho_1^2)(2\pi I_{0}(\kappa_1))^{-1}}{1+\rho_1^2-2\rho_1\cos{(\phi-\gamma_1)}} \exp\left[ \dfrac{\kappa_1 \{ \xi_1 \cos{(\phi-\eta_1)}-2\rho_1 \cos \nu_1 \}}{1+\rho_1^2-2\rho_1 \cos{(\phi-\gamma_1)}}  \right] \times\\ 
     \left[\frac{ (1-\rho_2^2)(C~2\pi I_{0}(\kappa_2))^{-1} \left (1+\nu\cos\theta \right)}{1+\rho_2^2-2\rho_2\cos{(\theta-\gamma_2)}}    \exp\left[ \dfrac{\kappa_2 \{ \xi_2 \cos{(\theta-\eta_2)}-2\rho_2 \cos \nu_2 \}}{1+\rho_2^2-2\rho_2 \cos{(\theta-\gamma_2)}}\right]  \right],
    \label{kj torus}
\end{multline}
where $0\leq \mu_j,\nu_j<2\pi$, and $0\leq \rho_j <1$, $\kappa_j>0$, and 
$\gamma_j=\mu_j+\nu_j$, $\xi_j=\sqrt{\rho_j^4+2\rho_j^2\cos{(2\nu_j)}+1}$, $\eta_j=\mu_j+\arg(\rho_j^2\cos{(2\nu_j)}+1+i\rho_j^2\sin{(2\nu_j)}), \mbox{~for~} j=1,2,$
which implies that

$$ h_{1}(\phi)= f_{kj}(\phi) \mbox{~and~}h^{*}_{2}(\theta)=\frac{1}{C}f_{kj}(\theta)\left(1+\nu\cos\theta \right),$$
 with a normalizing constant
$$C=\displaystyle \int_{0}^{2\pi} \left[ f_{kj}(\theta)~\left(1+\nu\cos\theta \right) \right]~d\theta.$$
Hence, the joint probability density function
in Equation (\ref{kj torus}) represents the Kato and Jones distribution on the surface of a curved torus.
Now let us consider $\mu=\pi/2, \nu_1=\pi, \kappa=1,\rho=0.3$, and $\nu=0.5$,
Figure-\ref{vm}(c) displays the histogram of the sampled values for the vertical angle $\theta$, generated using the proposed sampling method described in Section-\ref{ch:HAR}. The corresponding acceptance rates for $n=50000$ and partition $k=250$, under fixed values of $\rho_2$ and $\kappa_2$, are reported in Tables-\ref{table:kj rho fix tor HAR2} and \ref{table:kj kp fix tor HAR2}, respectively. These results demonstrate the efficiency of the proposed algorithm in generating samples from the target distribution of $\theta$.
\end{enumerate}

\begin{table}[h!]
\centering
\scalebox{0.9}{
\begin{tabular}{|l|c|c|c|c|c|c|c|c|c|c|}
\hline
$\kappa_2$ & 1 &2& 3& 4& 5 &6& 7 &8 &9 &10 \\
\hline
\hline
\textbf{Proposed } & 99.058 & 99.066& 99.110& 99.128 &99.098 &99.136 &99.130& 99.138 &99.144 &99.144\\
\hline
\end{tabular}}
\vspace{.1cm}
\caption{Acceptance percentage for the density  $\frac{1}{C}f_{kj}(\theta)\left(1+\nu\cos\theta \right)$ with  $\mu=\pi/2, \nu_1=\pi$, $\rho_2=0.5$  (fixed) and varying $\kappa_2$.}
\label{table:kj rho fix tor HAR2}
\end{table}

\begin{table}[h!]
\centering
\scalebox{1}{
\begin{tabular}{|l|c|c|c|c|c|c|c|c|c|c|}
\hline
$\rho_2$ & 0.1 &0.2& 0.3& 0.4& 0.5 &0.6& 0.7 &0.8 &0.9  \\
\hline
\hline
\textbf{Proposed } & 99.376& 99.274& 99.246& 98.834& 98.640& 98.290& 97.418& 96.216& 92.412 \\
\hline
\end{tabular}}
\vspace{.1cm}
\caption{Acceptance percentage for the density  $\frac{1}{C}f_{kj}(\theta)\left(1+\nu\cos\theta \right)$ with  $\mu=\pi/2, \nu_1=\pi$, $\kappa_2=1$  (fixed) and varying $\rho_2$.}
\label{table:kj kp fix tor HAR2}
\end{table}

\begin{figure}[h!]
\centering
\subfloat[ ]{%
{\includegraphics[trim=0 90 0 80 , width=0.35\textwidth,height=0.35\textwidth]{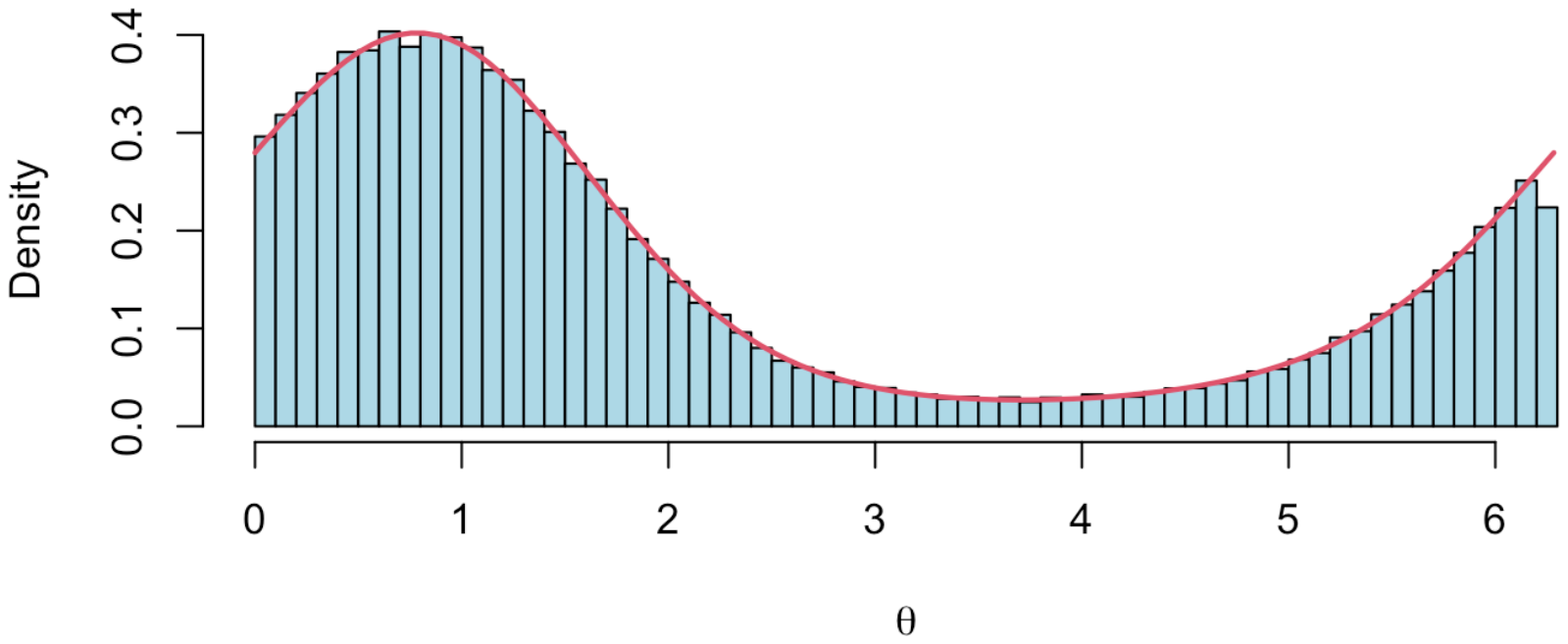}}}
\subfloat[]{%
{\includegraphics[trim=0 90 0 80 , width=0.35\textwidth,height=0.35\textwidth]{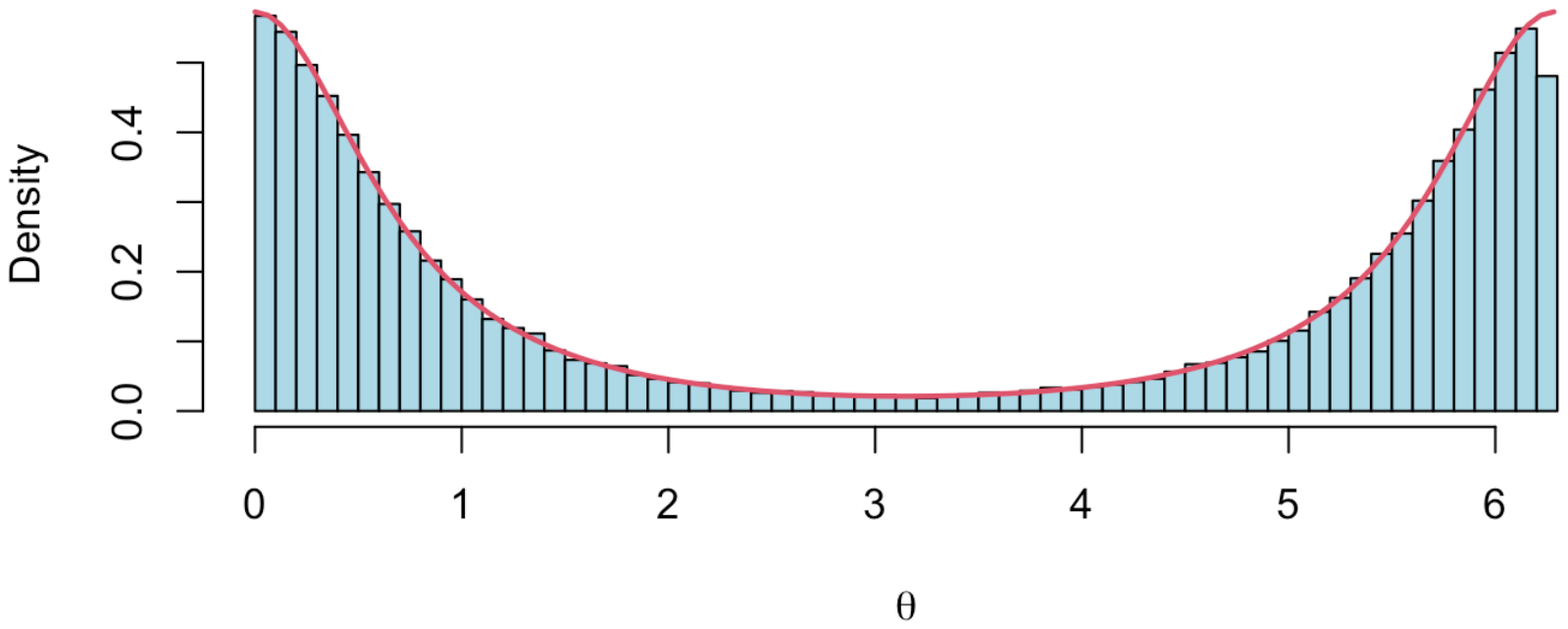}}}
\subfloat[]{%
{\includegraphics[trim=0 90 0 80 , width=0.35\textwidth,height=0.35\textwidth]{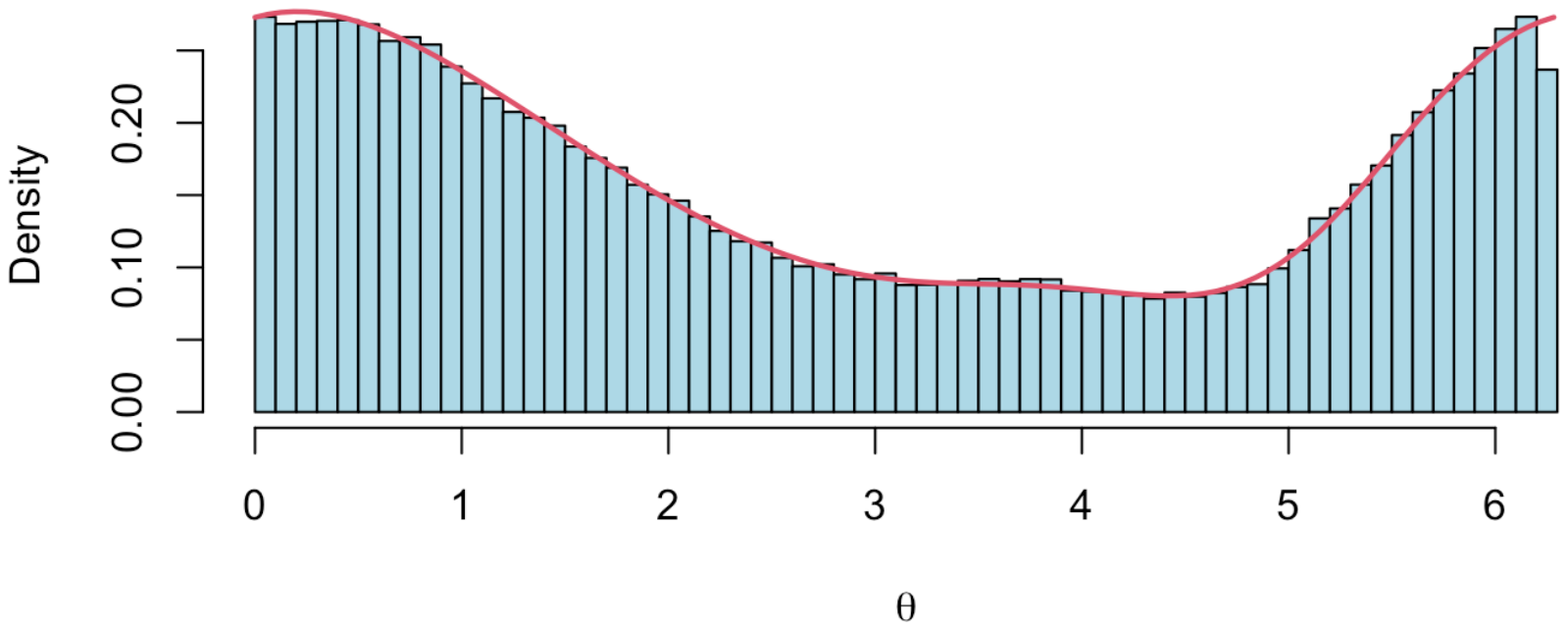}}}
\vspace{-0.3cm}
\caption{ (a) Histogram of the data from the density $h_2(\theta)$ with $\mu=\pi/3, \kappa=1$, and $\nu=0.5$ (b)  Histogram of the data from the density  $\frac{1}{C}f_{wc}(\theta)\left(1+\nu\cos\theta \right)$ with $\mu=0, \rho=0.5$, and $\nu=0.5$. (c) Histogram of the data from the  density  $\frac{1}{C}f_{kj}(\theta)\left(1+\nu\cos\theta \right)$ with $\mu=\pi/2, \nu_1=\pi, \kappa=1,\rho=0.3$, and $\nu=0.5$} \label{vm}
\end{figure}


\section{Some theoretical properties of the von Mises distribution on the surface of the curved torus} \label{voncos_properties}
The joint probability density function described in Equation (\ref{von torus}) characterizes the von Mises distribution on the surface of a curved torus. Thus, the density function 
typically exhibits asymmetry in $\theta$ but is symmetric in $\phi$. The asymmetry comes from the second part of the function, specifically due to the $\nu \cos \theta $ term. If $\nu=0$ or $\mu_2=0$, the density would become symmetric in both $\phi$ and $\theta$.  Figure-\ref{bv_voncos_desity_plot}(a) $\& $ (b) illustrate the bivariate symmetric density for $(\mu_1,\mu_2)=(0,0)$, while Figure-\ref{bv_voncos_desity_plot}(c) $\& $ (d) depict the bivariate asymmetric density for $(\mu_1,\mu_2)=(0,\pi/4)$. Throughout these representations, the parameters remain consistent: $(\kappa_1, \kappa_2)=(3,0.5)$ and $\nu=0.95$.

Many researchers have already studied von Mises distribution in great detail \cite[see][]{mardia2000directional, jammalamadaka2001topics}. But 
the marginal distribution, $h_{2}(\theta)$, is a new one on the circle where the density of the von Mises distribution is associated with the area element of the curved torus. So, in the following sections, we extensively studied the theoretical properties of the distribution of the vertical angle $\theta.$
The probability distribution function of the vertical angle $\theta$ is given in Equation (\ref{voncos}).
Some special cases can be noted as follows.
\begin{itemize}
    \item \textit{von Mises:} 
    As $\nu \xrightarrow{} 0$ the Equation (\ref{voncos}) becomes 
   $ h_2(\theta)=\frac{e^{\kappa\cos(\theta-\mu)}}{2\pi~I_{0}(\kappa)} $, which is the density of von Mises distribution.
   
     \item  \textit{Cardioid:}
   When $\kappa=0$, $I_{0}(0)=1 \mbox{~and~} I_{1}(0)=0$. Hence the Equation (\ref{voncos}) becomes 
 $ h_2(\theta)=\frac{1}{2\pi}\left(1+\nu\cos\theta \right)$ , which is the density of Cardioid distribution with location parameter at $\eta=0$  \cite[see][]{mardia2000directional}.
     
     \item \textit{Uniform:} 
 As $\nu \xrightarrow{} 0$, and $\kappa=0$, the Equation (\ref{voncos}) becomes 
 $ h_2(\theta)=\frac{1}{2\pi}$ , which is the density of circular uniform distribution.

  \item \textit{Dirac-delta distribution:} 
 As  $\kappa \rightarrow \infty$,  the model
in Equation (\ref{voncos}) represents a one point distribution for all $\nu$ with singularity at $\theta=\mu.$
\end{itemize}

Here, we further discuss the interpretation of the parameters by making some plots of probability density functions. Following Equation (\ref{voncos})  Figure-\ref{voncos_desity_plot}(a) is obtained by putting $\mu=\frac{\pi}{2}, \nu=0.5$, and varying $\kappa$ from $0$ to $4$ with each increment $1$. Note that for $\kappa=0$ it provides the plot of the Cardioid distribution.  Figure-\ref{voncos_desity_plot}(b) is obtained by putting $\mu=\frac{\pi}{2}, \kappa=1$, and varying $\nu$ from $0.1$ to $0.4$ in steps of $0.1$. We can see that from Figure-\ref{voncos_desity_plot}(a) $\&$ Figure-\ref{voncos_desity_plot}(b) that the density is asymmetric in general.  Figure-\ref{voncos_desity_plot}(c) is obtained by putting $\nu=0.5, \kappa=1$, and $\mu= 0, \frac{\pi}{3},\frac{2\pi}{3}, \pi.$ The Figure-\ref{voncos_desity_plot}(c) shows the interpretation of $\mu$, implying that the parameter $\mu$ has a crucial role in controlling the skewness of the model.

\begin{figure}[h!]
\centering
\subfloat[]{%
{\includegraphics[trim=20 20 20 20,width=0.35\textwidth,height=0.27\textwidth]{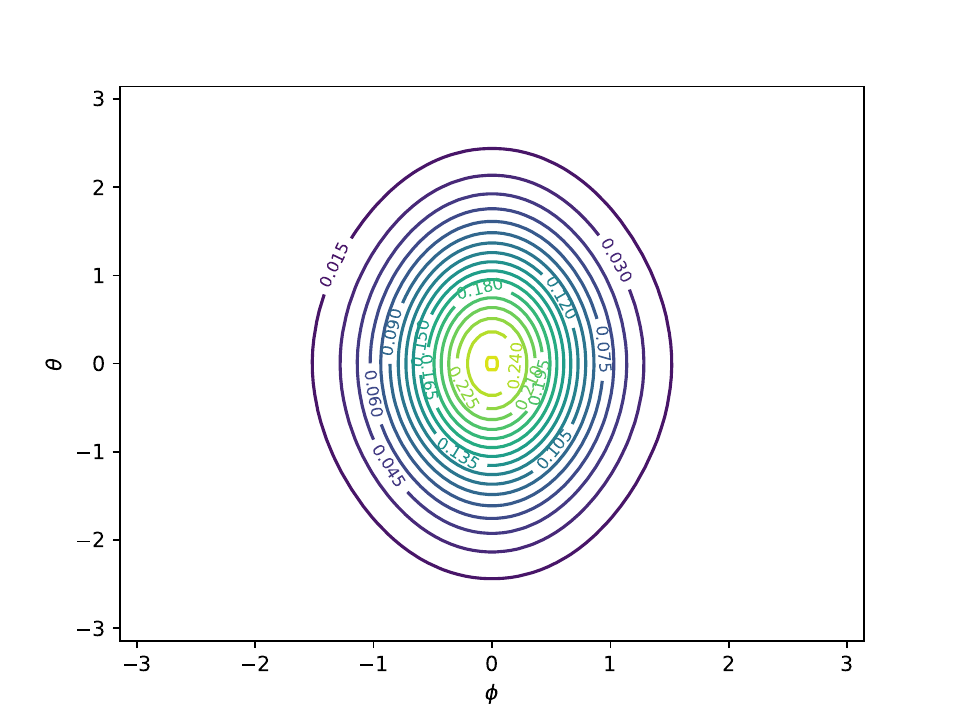}}}
\subfloat[]{%
{\includegraphics[trim=20 20 20 20,width=0.35\textwidth,height=0.27\textwidth]{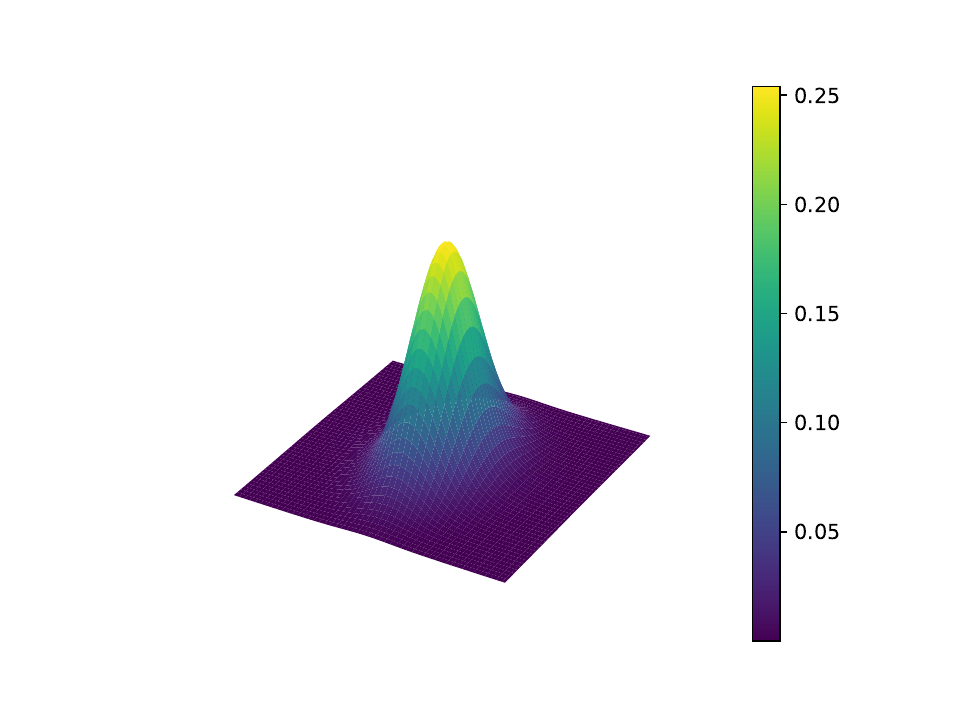}}}\hspace{2pt}
\subfloat[]{%
{\includegraphics[trim=20 20 20 20,width=0.35\textwidth,height=0.27\textwidth]{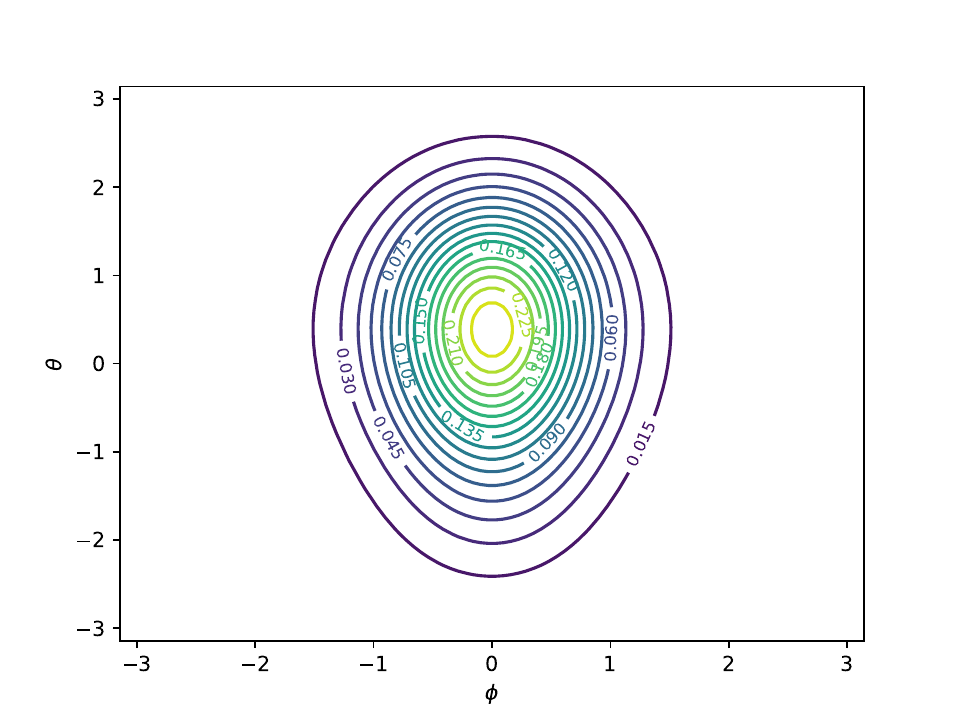}}}
\subfloat[]{%
{\includegraphics[trim=20 20 20 20,width=0.35\textwidth,height=0.27\textwidth]{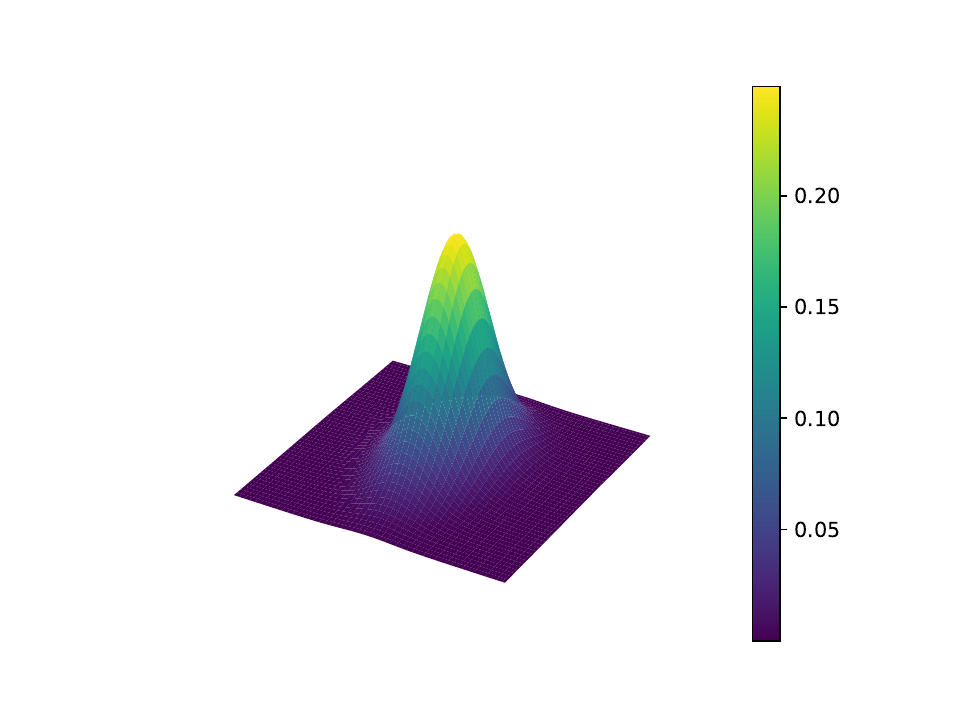}}}
\caption{ (a) contour plots and (b) 3D surface plots the proposed bivariate density function provided in Equation (\ref{von torus}) for $(\mu_1,\mu_2)=(0,0)$ \. Similarly, for $(\mu_1,\mu_2)=(0,\pi/4)$, we present (c) contour plot and (d) 3D surface plot. These visualizations maintain parameters throughout: $(\kappa_1, \kappa_2)=(3,0.5)$ and $\nu=0.95$.  For a better diagrammatic representation, the plots are shown in $[-\pi, \pi) \times [-\pi, \pi)$ instead of $[0, 2\pi) \times [0, 2\pi).$ } 
\label{bv_voncos_desity_plot}
\end{figure}

\begin{figure}[t!]
\centering
\subfloat[]{%
{\includegraphics[width=0.35\textwidth,height=0.27\textwidth]{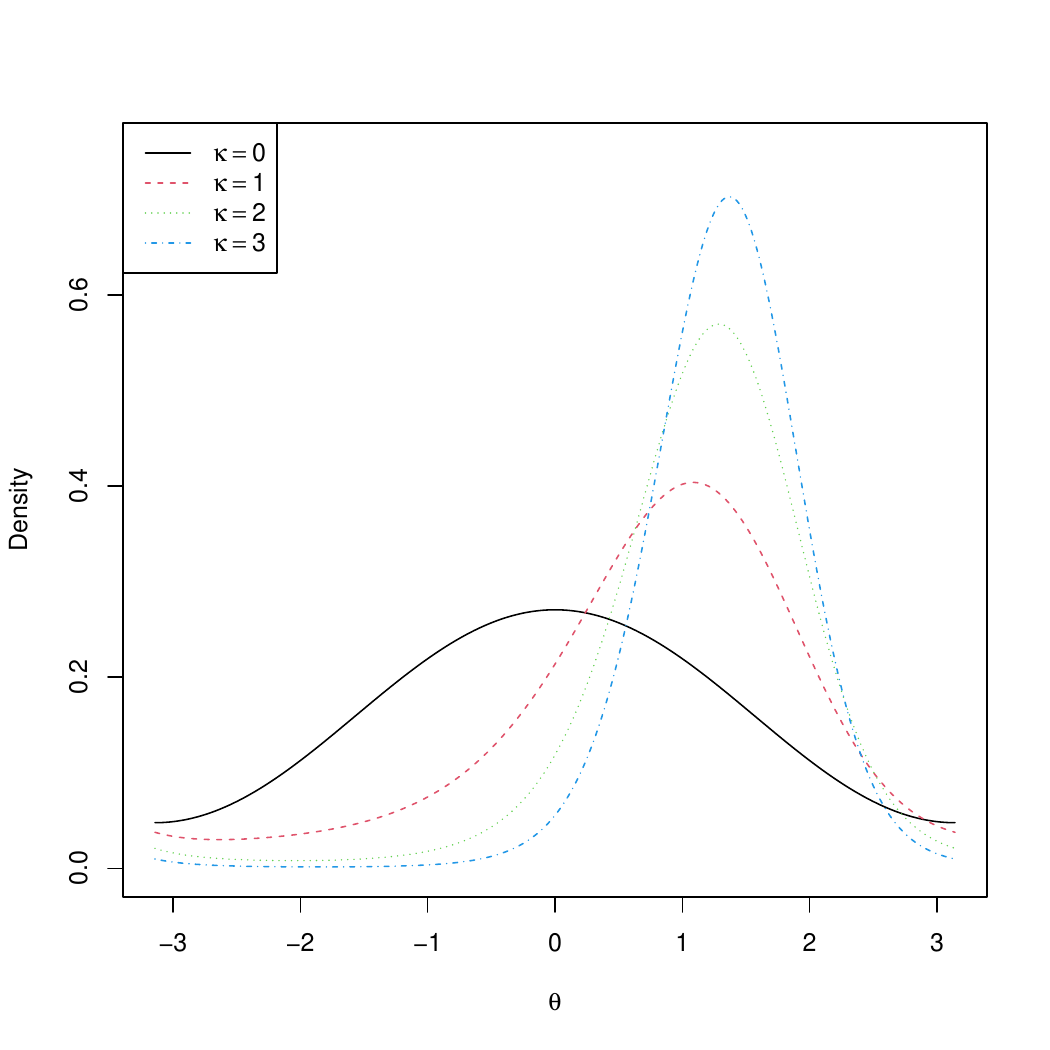}}}
\subfloat[]{%
{\includegraphics[width=0.35\textwidth,height=0.27\textwidth]{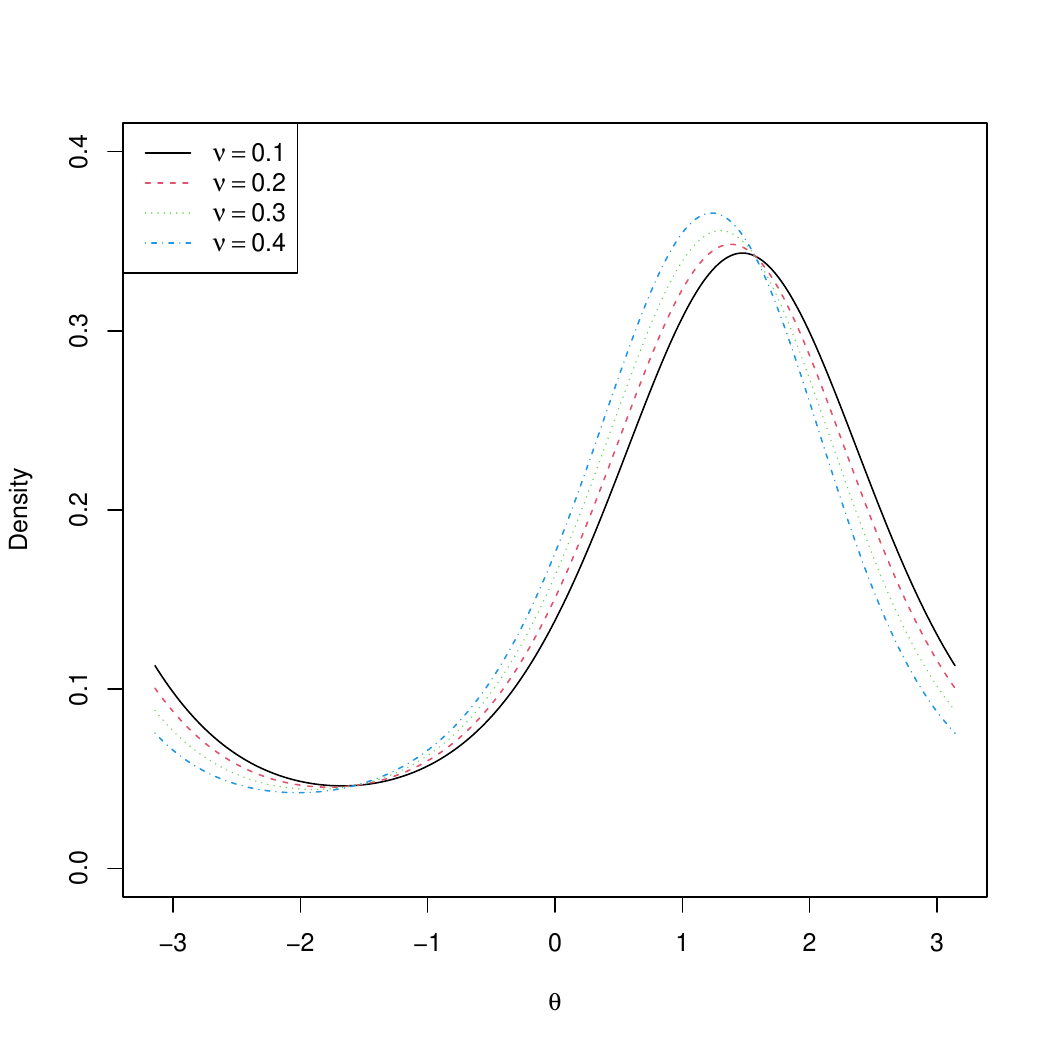}}}\hspace{2pt}
\subfloat[]{%
{\includegraphics[width=0.35\textwidth,height=0.27\textwidth]{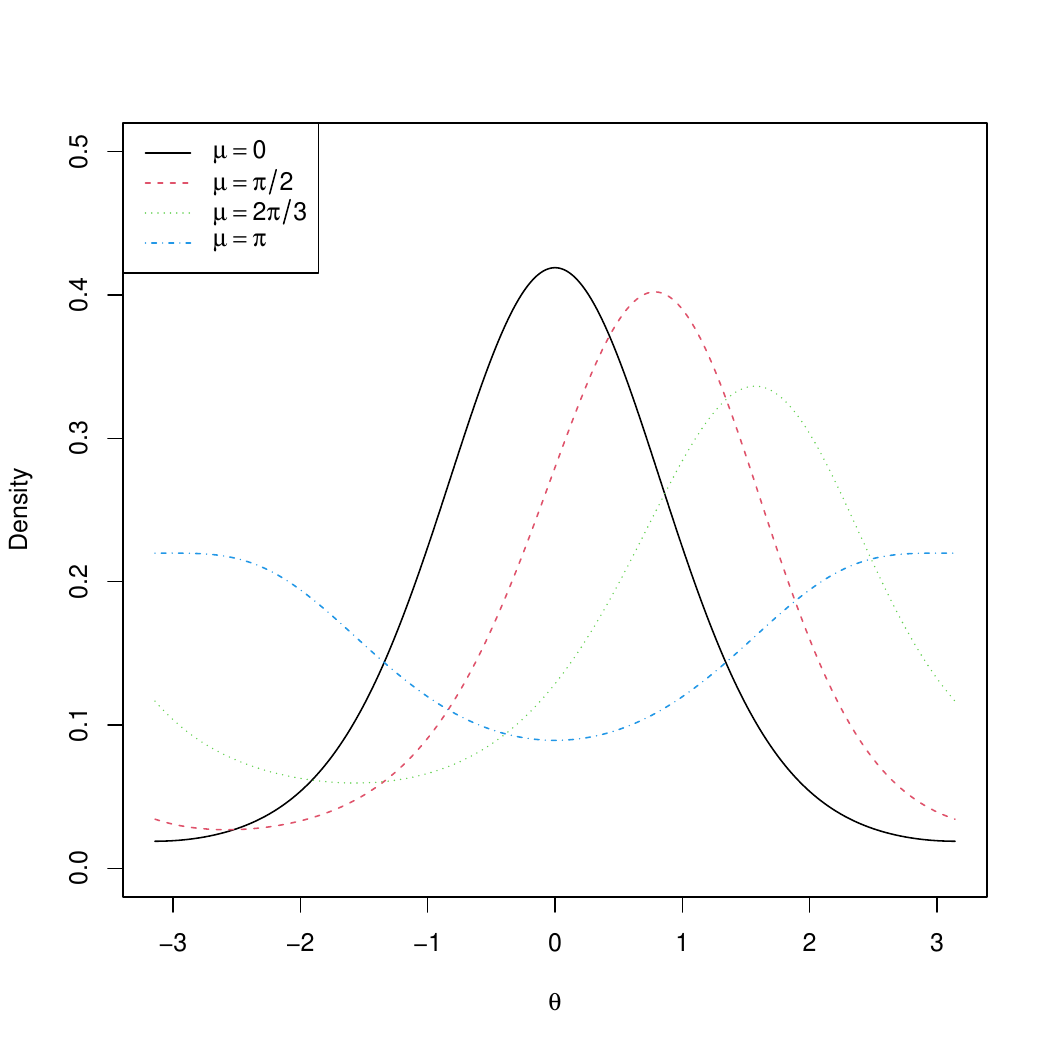}}}
\subfloat[]{%
{\includegraphics[width=0.35\textwidth,height=0.27\textwidth]{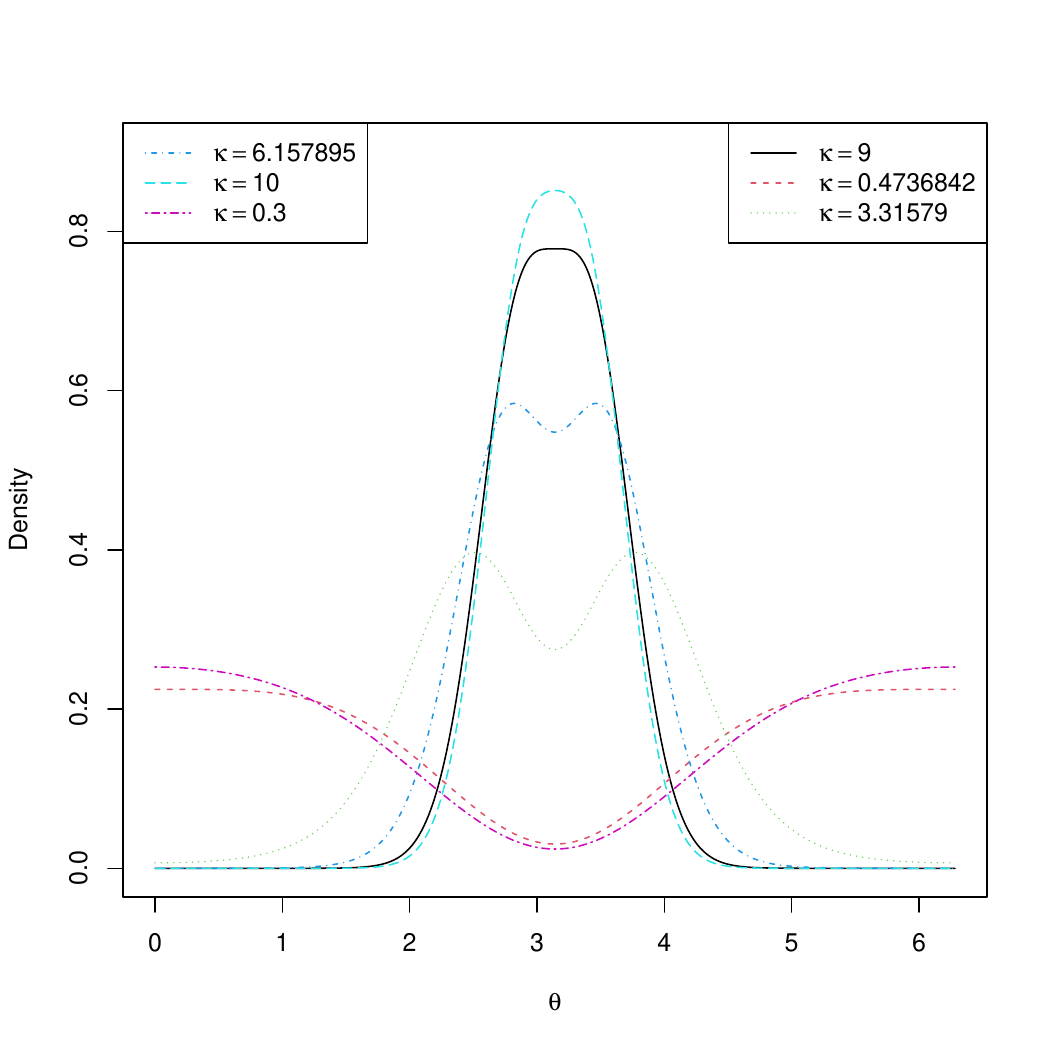}}}
\caption{ Density plot of the probability density function provided in Equation (\ref{voncos})  (a) $\mu=\frac{\pi}{2}$  $\nu=0.5$, along with  $\kappa=0,1,2,3,4$.  (b) $\mu=\frac{\pi}{2}$, $\kappa=1$,  along with $\nu=0.1,0.2,0.3,0.4$, (c)  $\nu=0.5,~~ \kappa=1$, along with $\mu= 0, \frac{\pi}{3}, \frac{2\pi}{3}, \pi.$ (d) $\mu=\pi,~~\nu=0.9$, and for different values of $\kappa$.  For a better diagrammatic representation, the plots in (a), (b), and (c) are shown in $[-\pi, \pi)$ instead of $[0, 2\pi).$ } 
\label{voncos_desity_plot}
\end{figure}

\subsection{The maximum entropy characterization}
\label{max_entropy}

The maximum entropy characterizations of the toroidal distributions described by Equation (\ref{voncos}) can be derived through an extension of the general result provided by \cite{kagan1973characterization} \cite[see][p. 409, Theorem 13.2.1]{}. This result has been applied to the von Mises distribution on the circle and the Fisher distribution on the sphere by \cite[pp. 65-66]{mardia1975statistics} and \cite[pp. 162-163]{rao1973linear}, respectively. 

The maximum entropy distribution on the circle is the von Mises distribution, having the probability density function in Equation (\ref{von mises}), and the maximum entropy distribution on the sphere is the Fisher distribution due to \cite{fisher1953dispersion}, having the probability density function. 

\begin{equation}
 f_{f}(\phi,\theta)=\frac{1}{2\pi}\frac{\kappa}{2 \sinh\kappa} e^{\kappa\cos\theta}\sin\theta,
 \label{fisher dist}
\end{equation}
where $\kappa>0, 0\leq \theta<\pi$, and $0\leq \phi<2\pi.$ It is important to note that the probability density function for the azimuthal angle (horizontal) and zenith angle (vertical angle) are 
$h_1(\phi)=\frac{1}{2\pi}$ and $ h_2(\theta)=\frac{\kappa}{2 \sinh\kappa} e^{\kappa\cos\theta}\sin\theta,$ respectively. Also, the Jacobian for the unit sphere is $J_s=\sin\theta$, which is associated with the angle  of vertical distribution. For a more detailed explanation of spherical distribution, the reader may refer to \cite{mardia2000directional} \cite[see][ch. 9]{}.

From the above discussion of the  Fisher distribution and its maximum entropy characterization, it can be observed that the distribution of the vertical angle plays a crucial role in the sphere. Similar can be observed for the toroidal distribution also. Here, we deduce the maximum entropy characterization of toroidal distribution in the following theorem.

\begin{theorem}
    
    The marginal density of  the maximum entropy distribution for the vertical angle $\theta$ on the curved torus  is given in Equation (\ref{voncos}), subject to the constraints,

    \begin{eqnarray}
        E(\cos\theta)&=&a_1\\
        E(\sin\theta)&=&a_2\\
E\left(\log(J_2(f(\phi,\theta)))\right)&\leq&a_3,
    \end{eqnarray}
where

    $       a_1=\frac{I_1(\kappa)}{I_0(\kappa)}\cos\mu ,~
        a_2=\frac{I_1(\kappa)}{I_0(\kappa)}\sin\mu, ~
        a_3=\log(\nu(1+\nu)), 
   $ and 
     $J_2f(\phi,\theta)=r(R+r\cos{\theta})=\nu(1+\nu \cos \theta) ~[\mbox{as we have assumed}~~~ \frac{r}{R}=\nu]$, is the Jacobian defined on Equation (\ref{jacobian}).
   \label{max_entropy_thm}
\end{theorem}

\begin{proof}

See the Appendix-\ref{max_entropy_proof} for the proof.
\end{proof}

\begin{rmk}
 If the distribution for the horizontal angle $\phi$ is the circular uniform distribution, $h_1(\phi)=\frac{1}{2\pi}$ for $0\leq \phi <2\pi,$ together with $h_2(\theta)$ as in Equation (\ref{voncos}), the maximum entropy distribution on curved torus is
$$h^{*}(\phi,\theta)= \frac{1}{2\pi} \left[ \frac{e^{\kappa\cos(\theta-\mu)}}{\left[2\pi(I_{0}(\kappa)+\nu \cos \mu~I_{1}(\kappa)) \right]} \left(1+\nu\cos\theta\right)\right], $$

where $0\leq \phi, \theta<2\pi$, $0\leq \mu<2\pi$, $\kappa>0$. Here, $h_1(\phi)=\frac{1}{2\pi}$ is the maximum entropy distribution without any restrictions.
\end{rmk}

\begin{rmk}
      If the distribution of the horizontal angle $\phi$ is the von Mises distribution in Equation (\ref{von mises}), together with $h_2(\theta)$ in Equation (\ref{voncos}), the maximum entropy distribution on curved torus is
$$  h^{*}(\phi,\theta)= \frac{e^{\kappa_1\cos(\phi-\mu_1)}}{2\pi I_{0}(\kappa_1)} \left[ \frac{e^{\kappa_2\cos(\theta-\mu_2)}}{\left[2\pi(I_{0}(\kappa_2)+\nu \cos \mu_2~I_{1}(\kappa_2)) \right]} \left(1+\nu\cos\theta \right)\right], $$

where $0\leq \phi, \theta<2\pi$, $0\leq \mu_1,\mu_2<2\pi$, $\kappa_1,\kappa_2>0$. Here, $h_1(\phi)=\frac{e^{\kappa_1\cos(\phi-\mu_1)}}{2\pi I_{0}(\kappa_1)}$ is the maximum entropy distribution subject to the  restrictions :

  \begin{eqnarray}
        E(\cos\phi)&=&\frac{I_1(\kappa_1)}{I_0(\kappa_1)}\cos\mu_1 ,~~~~
        E(\sin\phi)=\frac{I_1(\kappa_1)}{I_0(\kappa_1)}\sin\mu_1, \nonumber
    \end{eqnarray}

\end{rmk}


\subsection{Trigonometric Moments}
\label{trig_moments}

The trigonometric moments of the probability distribution in Equation (\ref{voncos}) are given by the following theorem:

\begin{theorem}
    If $\Theta \sim h_2(\theta)$ in Equation (\ref{voncos}) then the $p^{th}$ trigonometric moments  for $p=0,\pm 1, \pm2, \ldots$ are given by 

$$\Phi_{p}= \frac{\nu I_{p-1}(\kappa)e^{i(p-1)\mu}+2I_{p}(\kappa)e^{ip\mu}+\nu I_{p+1}(\kappa)e^{i(p+1)\mu}}{2\left[I_{0}(\kappa)+\nu \cos \mu~I_{1}(\kappa)\right]} $$
\label{ch.function}
\end{theorem}

\begin{proof}

See the Appendix-\ref{ch_fun_proof} for the proof.
\end{proof}

\begin{corr}
     As $\nu \rightarrow 0$ the $pth$ trigonometric moments for $p=0,\pm 1, \pm2, \ldots$ in Theorem-\ref{ch.function} becomes  
    $$ 
     \Phi_{p}=\frac{I_{p}(\kappa)e^{ip\mu}}{I_{0}(\kappa)}      ,$$ which is the  $pth$ trigonometric moments for von Mises distribution.
\end{corr}

\begin{corr}
      When $\kappa=0$ the $pth$ trigonometric moments for $p=0,\pm 1, \pm2, \ldots$ in Theorem-\ref{ch.function} becomes  
    $$\Phi_{p}=\frac{\nu I_{p-1}(0)e^{i(p-1)\mu}+2~I_{p}(0)e^{ip\mu}+\nu   I_{p+1}(0)e^{i(p+1)\mu}}{2\left[I_{0}(0)+\nu \cos \mu~I_{1}(0)\right]}     ,$$ as $I_{p+1}(0)=I_{p-1}(0)=0, \mbox{~for~}p \neq \pm 1$ $\mbox{~and~} I_{0}(0)=1 ,$ then we can write trigonometric moments as 

    $$  \Phi(\theta)=\frac{\nu}{2}   ,$$
    which is the  $pth$ trigonometric moments for Cardioid distribution with location parameter $\eta=0$, for more details \cite[see][]{jammalamadaka2001topics}.
\end{corr}

\subsection{Condition of Symmetry}
\label{condition_symm}

The density in Equation (\ref{voncos}) is generally an asymmetric distribution. Hence, the condition for the symmetry is given in the following corollary

\begin{corr}
     The density in Equation (\ref{voncos}) is symmetric iff $\mu=0$, $\kappa=0$ or $\nu\xrightarrow{}0$.
    \label{symm_con}
\end{corr}

\begin{proof}
    See Appendix-\ref{symm_con_proof}, for the proof.
\end{proof}

 Of course, the symmetric special cases with  $\nu\xrightarrow{}0$ and $\kappa=0$  are the von Mises and Cardioid distributions; the wider two-parameters family of symmetric distributions arising when $\mu=0$  as discussed in Section \ref{symm_case}.

\subsection{Condition of Modality}

The probability density function in Equation (\ref{voncos}) can be both unimodal or multimodal depending on the parameters
$\mu$, $\kappa$ and $\nu$ . In this subsection, we give necessary and sufficient conditions for unimodality. Here, the proposed probability density function is differentiable with respect to $\theta\in [0,2\pi).$ Now, the derivative of Equation (\ref{voncos})  with respect to $\theta$ is 
\begin{eqnarray}
     \frac{d}{d\theta}h_2(\theta)&\propto& \frac{d}{d\theta} \left[ e^{\kappa\cos{(\theta-\mu)}}\left(1+\nu\cos{(\theta)} \right) \right] \nonumber\\
       &=& -{\kappa}\mathrm{e}^{{\kappa}\cos\left({\theta}-{\mu}\right)}\cdot\left(1+\nu\cos\left({\theta}\right)\right)\sin\left({\theta}-{\mu}\right)-\nu\mathrm{e}^{{\kappa}\cos\left({\theta}-{\mu}\right)}\cdot\sin\left({\theta}\right)
\end{eqnarray}
To identify the number of modes, we need to solve the following equation $\frac{d}{d\theta}h_2(\theta)= 0 $, which implies that
\begin{equation}
        -{\kappa}\mathrm{e}^{{\kappa}\cos\left({\theta}-{\mu}\right)}\cdot\left(1+\nu\cos\left({\theta}\right)\right)\sin\left({\theta}-{\mu}\right)-\nu\mathrm{e}^{{\kappa}\cos\left({\theta}-{\mu}\right)}\cdot\sin\left({\theta}\right)=0
\end{equation}
simplifying to

\begin{equation}
    b_1 \left[\sin{\theta}+\nu\cos{\theta} \sin{\theta}\right]-b_2\left[ \cos{\theta}+\nu \cos^2{\theta} \right]+b_3 \sin{\theta=0},
    \label{unimodal critical eq}
\end{equation}

where, $b_1=\cos{\mu}, b_2=\sin{\mu}, \mbox{~and~} b_3=\frac{\nu}{\kappa}$

As discussed in a related context by \cite{yfantis1982extension},  this  equation  is  equivalent  to  four degree  equation which can be obtained by the transformation $x=\tan(\theta/2)$, so we can use 
$$\sin\theta=\frac{2x}{1+x^2}, \mbox{~and~} \cos\theta=\frac{1-x^2}{1+x^2}$$ in the Equation (\ref{unimodal critical eq}), hence it becomes 
\begin{equation}
    d_4x^4+d_3x^3+d_2x^2+d_1x+d_0=0,
    \label{unimodal quartic}
\end{equation}

where $d_4=b_2(1-\nu), d_3=2b_3+2b_1(1-\nu), d_2=2b_2\nu, d_1=2b_3+2b_1(1+\nu) \mbox{~and~} d_0=-b_2(1
+\nu).$
This quartic equation can be solved using  Ferrari’s method,  as discussed by \cite{uspensky1948theory}. The discriminant of this equation is 
\begin{eqnarray}
    \Delta&=& 
    d_{0}^2\left[ -27d_{3}^4+144d_{2}d_{3}^2d_{4}-128d_{2}^2d_{4}^2-192d_{1}d_{3}d_{4}^2   \right]\nonumber\\
    &+&d_{0}\left[ -4d_{2}^3d_{3}^2+18d_{1}d_{2}d_{3}^3+16d_{2}^4d_{4}-80d_{1}d_{2}^2d_{3}d_{4}-6d_{1}^2d_{3}^2d_{4}+144d_{1}^2d_{2}d_{4}^2 \right] \nonumber\\
    &+&d_{1}^2d_{2}^2d_{3}^2-4d_{1}^3d_{3}^3-4d_{1}^2d_{2}^3d_{4}+18d_{1}^3d_{2}d_{3}d_{4}-27d_{1}^4d_{4}^2+256d_{0}^3d_{4}^3
    \label{discriminant}
\end{eqnarray}
Equation (\ref{unimodal quartic}) is known to have four real roots or four complex roots if $\Delta>0$,  and two real roots and two complex roots if $\Delta<0$. This immediately ensures that the density has no more than two modes.  Moreover,  the distribution is bimodal when $\Delta>0$ and unimodal when $\Delta<0$. 
 Because $d_i, i=1,\cdots,4$, are functions of $\mu,\kappa, \mbox{~and~} \nu$ the conditions for unimodality can in principle be written out in terms of these three parameters.

\begin{rmk}
    When $\mu=0$ then $b_1=1,b_2=0$, and $b_3=\frac{\nu}{k}$. Eventually  $d_4=d_2=d_0=0, d_3=\frac{2\nu}{\kappa}+2(1-\nu)$, and $d_1=  \frac{2\nu}{\kappa}+2(1+\nu)$. So, the Equation (\ref{unimodal quartic}) becomes 

    \begin{equation}
        d_3x^3+d_1x=0
    \end{equation}

with the discriminant $\Delta=-4d_3d_{1}^3$. Now, we have  $\kappa>0$, and $\nu\in (0,1)$, hence $d_1,d_3$ are always positive and as a consequence, the discriminant $\Delta=-4d_3d_{1}^3$ is always negative. Thus, the probability density function in Equation (\ref{voncos}) is unimodal for $\mu=0.$  The Figure-\ref{voncos_desity_plot}(c) also agrees with this.
\label{rmk_uni_mod}
\end{rmk}

\begin{rmk}
   When $\mu=\pi$ then $b_1=-1,b_2=0$, and $b_3=\frac{\nu}{\kappa}$. Eventually  $d_4=d_2=d_0=0, d_3=\frac{2\nu}{\kappa}-2(1-\nu)$, and $d_1=\frac{2\nu}{\kappa}-2(1+\nu)$, so, $d_1, d_2$ can be positive or negative depending upon the values of $\kappa >0$, and $\nu\in (0,1)$.
    
    Similarly, as above, we can have the following discriminant $\Delta=-4d_3d_{1}^3$. The bimodality and unimodality depend on the sign of the $d_1, d_3.$ Let us consider the following cases:

    \begin{itemize}
        \item Case-1: when $d_1,d_3<0$, we have $\Delta<0$ which implies unimodality. Now, $d_1<0 \implies \kappa> \frac{\nu}{1+\nu}$, and $d_3<0 \implies \kappa> \frac{\nu}{1-\nu}$. Together it gives the unimodality if $\kappa> \frac{\nu}{1-\nu}$.

         \item Case-2: when $d_1,d_3>0$, we have $\Delta<0$ which implies unimodality. Now, $d_1>0 \implies \kappa< \frac{\nu}{1+\nu}$, and $d_3>0 \implies \kappa<\frac{\nu}{1-\nu}$. Together it gives the unimodality if $\kappa< \frac{\nu}{1+\nu}$.

         \item Case-3: when $d_1<0$ and $d_3>0$, we have $\Delta>0$ which implies bimodality. Now, $d_1<0 \implies \kappa> \frac{\nu}{1+\nu}$, and $d_3>0 \implies \kappa<\frac{\nu}{1-\nu}$. Together, it gives the bimodality if 
         $\frac{\nu}{1+\nu}<\kappa< \frac{\nu}{1-\nu}$.

          \item Case-4: when $d_1>0$ and $d_3<0$, we have $\Delta>0$ which implies bimodality. Now, $d_1>0 \implies \kappa< \frac{\nu}{1+\nu}$, and $d_3<0 \implies \kappa>\frac{\nu}{1-\nu}$. Since $\nu\in (0,1)$, this case is not possible.
    \end{itemize}
   In particular, we have chosen $\nu=0.9$, so $\frac{\nu}{1+\nu}=0.4736842$, and $\frac{\nu}{1-\nu}=9$. Therefore, from Figure-\ref{voncos_desity_plot}(d), we can see that when $\kappa=9,10,0.4736842$, and $0.3$, then the probability density function in Equation (\ref{voncos}) is unimodal, which agrees with Case-1 and Case-2.When $0.4736842<\kappa<9$, that is $\kappa=3.3157895, 6.1578947$,  then the probability density function in Equation (\ref{voncos}) is bimodal, which agrees with Case-3.
\end{rmk}

 \subsection{Divergence From the Cardioid Distribution}
 \label{div_from_cardiod}
In this section, we compare the probability density function as given in Equation (\ref{voncos})
with the probability density function of the Cardioid distribution 

\begin{equation}
    f_{c}(\theta)=\frac{1}{2\pi}\left(1+\nu\cos\theta \right)
    \label{cardioid_dist}
\end{equation}
where $\nu\in (0,1).$
Calculating the KL divergence between $h_2$ and $f_c$ we get

\begin{eqnarray}
    D_{KL}(f_c,h_2)&=&\log((I_{0}(\kappa)+\nu\cos \mu~I_{1}(\kappa)))-\frac{\nu\kappa~\cos \mu}{2},
        \label{divergence_eqa}
\end{eqnarray}
see \ref{kl_appendix} in Appendix for details.
Clearly $D_{KL}(f_c,h_2)=0$ for $\kappa=0$. To understand the sensitivity of  $D_{KL}$ with respect to $\kappa$, we consider 

\begin{eqnarray}
    \frac{\partial D_{KL}(f_c,h_2) }{\partial \kappa}&=&\frac{I_{1}(\kappa)+\nu\cos \mu~[I_{0}(\kappa)+I_{2}(\kappa)]}{(I_{0}(\kappa)+\nu \cos \mu~I_{1}(\kappa))}-\frac{\nu \cos \mu}{2}\nonumber\\
    &=&\frac{A(\kappa)+\nu\cos \mu~[1+1-\frac{2A(\kappa)}{\kappa}]}{(1+\nu \cos \mu~A(\kappa))}-\frac{\nu \cos \mu}{2},\nonumber\\
    \nonumber
\end{eqnarray}
where $A(\kappa)=\frac{I_1(\kappa)}{I_0(\kappa)}$ and as $\kappa \rightarrow \infty$ then $A(\kappa) \rightarrow 1$ \cite[see][Section 4.2.1]{jammalamadaka2001topics}, which leads us to the following equation

\begin{eqnarray}
 \lim_{\kappa\rightarrow \infty}   \frac{\partial D_{KL}(f_c,h_2) }{\partial \kappa}
    &=&\frac{1+2\nu\cos \mu~}{1+\nu \cos \mu}-\frac{\nu \cos \mu}{2}=\frac{2+3\nu\cos \mu-\nu^2\cos^2 \mu~}{1+\nu \cos \mu}.
    \label{KL_cal}
\end{eqnarray}

\subsection{Maximum Likelihood Estimation}
\label{MLE}

Let $\theta_1,\cdots, \theta_n$ be the set of i.i.d observations from the probability density function given in Equation (\ref{voncos}), with the parameters $\kappa,\mu$, and $\nu$. The log-likelihood function is given by 

\begin{eqnarray}
    l&=&\sum_{i=1}^n\kappa \cos(\theta_i-\mu)+\sum_{i=1}^n \log \left[(1+\nu \cos\theta_i)\right]
    -n \log(2\pi)-n\log(I_0(\kappa)+\nu\cos\mu I_1(\kappa))\nonumber\\
    \label{log_liklihood}
\end{eqnarray}
We use the identities $ \frac{dI_0(\kappa)}{d\kappa}=I_1(\kappa), \frac{dI_1(\kappa)}{d\kappa}=\frac{I_0(\kappa)+I_2(\kappa)}{2}, A(\kappa)=\frac{I_1(\kappa)}{I_0(\kappa)},~ \frac{I_2(\kappa)}{I_0(\kappa)}=1-\frac{2A(\kappa)}{\kappa}.$  \cite[see][pp. 40]{mardia2000directional}\\ 

Differentiating Equation (\ref{log_liklihood})  with respect to $\mu$, $\kappa$, and $\nu$, respectively and equating with zero we get the following equations

\begin{eqnarray}
    \frac{\partial l}{\partial \mu}&=&\sum_{i=1}^n\kappa \sin(\theta_i-\mu)+\frac{n\nu  A(\kappa)\sin \mu}{1+\nu\cos\mu A(\kappa)}=0
    \label{mulikeli}
\end{eqnarray}

\begin{eqnarray}
   \frac{\partial l}{\partial \kappa}=\sum_{i=1}^n\cos(\theta_i-\mu)-n\frac{A(\kappa)+\nu\cos \mu~[1-\frac{A(\kappa)}{\kappa}]}{(I_{0}(1+\nu \cos \mu~A(\kappa))}&=&0
    \label{kappalikeli}
\end{eqnarray}

\begin{eqnarray}
 \frac{\partial l}{\partial \nu}=\sum_{i=1}^n \frac{\cos\theta_i}{[(1+\nu \cos\theta_i)]} -n\frac{A(\kappa)\cos \mu}{(1+\nu \cos \mu~A(\kappa))}&=&0.
    \label{alikeli}
\end{eqnarray}

As the closed-form solution of the above equations is intractable, one can obtain the numerical ones.

\textbf{Observed information matrix:}
To obtain the observed information matrix we consider the negative of the second
 derivatives of $l$ with respect to  $\mu,\kappa$, and $\nu$  as following:
\begin{eqnarray}
   J_{\mu\mu} =-\frac{\partial^2 l}{\partial \mu^2}&=&-\frac{n\nu  A(\kappa)(\nu A(\kappa)+\cos \mu)}{1+\nu\cos\mu A(\kappa)}
    \nonumber
\end{eqnarray}

\begin{eqnarray}
   J_{\mu\kappa} =- \frac{\partial l}{\partial \mu \partial \kappa}&=&-\sum_{i=1}^n \sin(\theta_i-\mu)-\frac{n\nu  A^{'}(\kappa)\sin \mu}{(1+\nu\cos\mu A(\kappa))^2}
   \nonumber
\end{eqnarray}

\begin{eqnarray}
   J_{\mu\nu} = -\frac{\partial l}{\partial \mu \partial \nu}&=&-\frac{n A(\kappa)\sin \mu}{(1+\nu\cos\mu A(\kappa))^2}
   \nonumber
\end{eqnarray}

\begin{eqnarray}
   J_{\kappa\kappa} = -\frac{\partial^2 l}{\partial \kappa^2} &=& \frac{n}{(1+\nu\cos\mu A(\kappa))^2} \bigg[ (1+\nu\cos\mu A(\kappa))\left(A'(\kappa) - \nu \cos \mu \left(\frac{\nu A'(\kappa)-A(\kappa)}{\kappa^2} \right) \right) \nonumber \\
    & &\hspace{2cm} -   \nu A'(\kappa) \cos \mu  \left (A(\kappa)+\nu\cos \mu~ \left[1-\frac{A(\kappa)}{\kappa}\right]\right)  \bigg]
     \nonumber
\end{eqnarray}

\begin{eqnarray}
    J_{\nu\nu} =-\frac{\partial^2 l}{\partial \nu^2}&=& --\frac{nA^2(\kappa)\cos^2 \mu}{(1+\nu \cos \mu~A(\kappa))^2}+\sum_{i=1}^n \frac{\cos^2\theta_i}{[(1+\nu \cos\theta_i)]^2} 
    \nonumber
\end{eqnarray}

\begin{eqnarray}
  J_{\kappa\nu} = - \frac{\partial l}{\partial \kappa \partial \nu}&=&n\frac{\cos \mu~ \left[1-\frac{A(\kappa)}{\kappa}\right]-A^2(\kappa) \cos \mu}{(1+\nu\cos\mu A(\kappa))^2}
   \nonumber
\end{eqnarray}

\textbf{Expected information matrix} Elements of the expected information matrix follow those of the observed information matrix. Denote $\frac{1}{n}$ times the expected information matrix by $\iota.$ Then we have the following:

\begin{eqnarray}
   \iota_{\mu\mu} =\frac{1}{n}E \left[-\frac{\partial^2 l}{\partial \mu^2} \right]&=&-\frac{\nu  A(\kappa)(\nu A(\kappa)+\cos \mu)}{1+\nu\cos\mu A(\kappa)}
    \nonumber
\end{eqnarray}

\begin{eqnarray}
   \iota_{\mu\kappa} &=&  \int_{0}^{2\pi} \sin(\theta-\mu) h_2(\theta) d\theta-\frac{\nu  A^{'}(\kappa)\sin \mu}{(1+\nu\cos\mu A(\kappa))^2}
   \nonumber
\end{eqnarray}

\begin{eqnarray}
   \iota_{\mu\nu}&=&-\frac{ A(\kappa)\sin \mu}{(1+\nu\cos\mu A(\kappa))^2}
   \nonumber
\end{eqnarray}

\begin{eqnarray}
   \iota_{\kappa\kappa} &=& \frac{1}{(1+\nu\cos\mu A(\kappa))^2} \bigg[ (1+\nu\cos\mu A(\kappa))\left(A'(\kappa) - \nu \cos \mu \left(\frac{\nu A'(\kappa)-A(\kappa)}{\kappa^2} \right) \right) \nonumber \\
    & &\hspace{2cm} -   \nu A'(\kappa) \cos \mu  \left (A(\kappa)+\nu\cos \mu~ \left[1-\frac{A(\kappa)}{\kappa}\right]\right)  \bigg]
     \nonumber
\end{eqnarray}

\begin{eqnarray}
    \iota_{\nu\nu}&=& \frac{A^2(\kappa)\cos^2 \mu}{(1+\nu \cos \mu~A(\kappa))^2}- \int_{0}^{2\pi}
 \frac{\cos^2\theta}{[(1+\nu \cos\theta)]^2} h_2(\theta) d\theta.
    \nonumber
\end{eqnarray}

\begin{eqnarray}
  \iota_{\kappa\nu} &=&\frac{\cos \mu~ \left[1-\frac{A(\kappa)}{\kappa}\right]-A^2(\kappa) \cos \mu}{(1+\nu\cos\mu A(\kappa))^2}
   \nonumber
\end{eqnarray}

\section{Special case: Two-parameter symmetric and unimodal}
\label{symm_case}
In this section, we study the properties of symmetric and unimodal cases of the probability density function given in Equation (\ref{voncos}) in detail.

\subsection{Probability density function}
\label{pdf_symm_sec}
From Theorem-\ref{symm_con} and Remark-\ref{rmk_uni_mod}  we observe that $\mu=0$ implies the model in Equation (\ref{voncos}) is the symmetric and uni-model. Hence, the symmetric unimodal sub-model is 

    \begin{equation}
    h_3(\theta)=\frac{e^{\kappa\cos\theta}\left(1+\nu\cos\theta \right)}{2\pi(I_{0}(\kappa)+\nu~I_{1}(\kappa))},
    \label{symm_voncos}
\end{equation}
where, $\theta \in[0,2\pi)$, $\kappa>0$, and $\nu\in (0,1).$ This above density also has similar special cases as of the density $h_2(\theta)$ in Equation (\ref{voncos}) that is von Mises as $\nu \xrightarrow{} 0$, Cardioid for $\kappa=0$, Uniform as $\nu \xrightarrow{} 0$, and $\kappa=0$, and Dirac-delta distribution as  $\kappa \rightarrow \infty$ for all $\nu$ with singularity at $\theta=0$. Now, in the following, we discuss some of the properties of the probability density function in Equation (\ref{symm_voncos}). 

\subsection{Moments, mode, and anti-mode}
\label{moment_mode}
For the symmetric and unimodal case, we have $\mu=0$. In this case  the $pth$ trigonometric moments for $p=0,\pm 1, \pm2, \ldots$ in Theorem-\ref{ch.function} becomes  
    $$\Phi_{p}=\frac{\nu I_{p-1}(\kappa)+2~I_{p}(\kappa)+\nu   I_{p+1}(\kappa)}{2\left[I_{0}(\kappa)+\nu ~I_{1}(\kappa)\right]}.$$

    Now, for $p=1$ we can write it as follows \cite[see][pp. 27-28]{jammalamadaka2001topics}

    $$\Phi_{1}=\rho_1e^{i\mu_1},$$
    where, 
    $$\rho_1=\left[\frac{\nu I_{0}(\kappa)+2~I_{1}(\kappa)+\nu   I_{2}(\kappa)}{2\left[I_{0}(\kappa)+\nu ~I_{1}(\kappa)\right]} \right], \mbox{~and~} \mu_1=0 $$
   are the mean resultant length and the mean direction of the probability density function in Equation (\ref{symm_voncos}).

 The circular variance, denoted by $v$, is a common measure of variation on the circle, where $0 < v < 1$. It is calculated as $v = 1-\rho_1$, where $\rho_1$ represents a measure of concentration. In the given contour plot (Figure-\ref{circular_variance_kl}(a)), the numerical values of $v$ are depicted as functions of $\rho_1$ and $\kappa$. The maximum value of circular variance is $1$, which occurs in the case of a uniform distribution.

From the plot, it's evident that as $\kappa$ approaches $0$ and $\nu$ approaches $0$, the values of $v$ tend to get closer to $1$, aligning with the behavior expected for a uniform distribution, which is a special case of the symmetric model represented by Equation (\ref{symm_voncos}). Furthermore, as $\kappa$ increases, circular variance decreases for all values of $\nu$, and vice versa. 

As $\nu$ approaches $0$, it can be shown that $v = 1 - A(\kappa)$, where $A(\kappa) = \frac{I_1(\kappa)}{I_0(\kappa)}$ for all $\kappa$. When $\kappa$ equals $0$, $v = \frac{\nu}{2}$ for all $\nu$. Finally, as $\kappa$ tends towards infinity, $v$ approaches $0$ for all $\nu$. A similar contour plot emerges when considering another measure of variation on the circle, namely the circular dispersion introduced by \cite{fisher1993statistical}.\\

 Since the maximum value of $\cos \theta=1$ at $\theta=0$, then the probability density function Equation (\ref{symm_voncos}) is maximum at $\theta=0$ i.e., is the modal direction having the maximum value  
    $$f(0)=\frac{e^{\kappa}\left(1+\nu \right)}{2\pi(I_{0}(\kappa)+\nu~I_{1}(\kappa))}.$$
Again  since the minimum value of $\cos \theta=-1$ at $\theta=\pm \pi$, then the probability density function in Equation (\ref{symm_voncos}) is minimum at $\theta=\pm \pi$ i.e., is the anti-modal direction having the minimum value  
    $$f(\pm \pi)=\frac{e^{-\kappa}\left(1-\nu \right)}{2\pi(I_{0}(\kappa)+\nu~I_{1}(\kappa))}.$$

\subsection{KL divergence and Maximum Likelihood Estimation:}
\label{kl_mle_symm}
 In this case, we put $\mu=0$ in   Equation (\ref{divergence}) and Equation (\ref{KL_cal})  which yields 

\begin{eqnarray}
    D_{KL}(f_c,h_3)
        &=& \log((I_{0}(\kappa)+\nu~I_{1}(\kappa)))-\frac{\nu\kappa}{2}, 
        \label{symm_kl}
\end{eqnarray}
     and
     \begin{eqnarray}
    \frac{\partial D_{KL}(f_c,h_3) }{\partial \kappa}
    &=&\frac{2+3\nu-\nu^2~}{1+\nu} >0, 
 \label{symm_deri_kl}
\end{eqnarray}
as $\nu \in (0,1).$ Figure-\ref{circular_variance_kl}(b) \& Figure-\ref{circular_variance_kl}(c) also agrees with the calculation. Hence, as $\kappa$ increases, the divergence between the Cardioid distribution and the proposed symmetric and unimodal model 
 increases.\\

\begin{figure}[h!]
\centering
\subfloat[]{%
{\includegraphics[width=0.6\textwidth,height=0.4\textwidth]{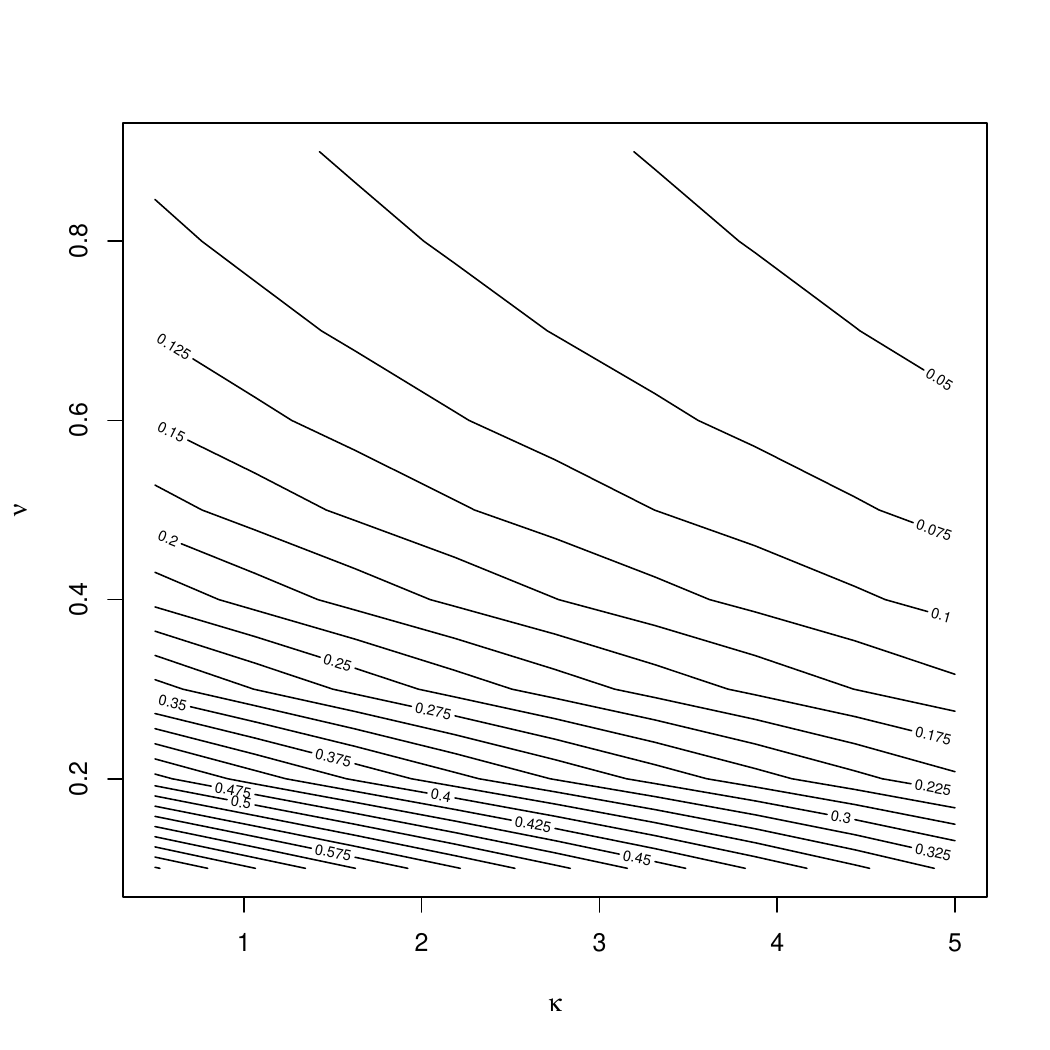}}}\hspace{2pt}
\subfloat[]{%
{\includegraphics[width=0.4\textwidth,height=0.4\textwidth]{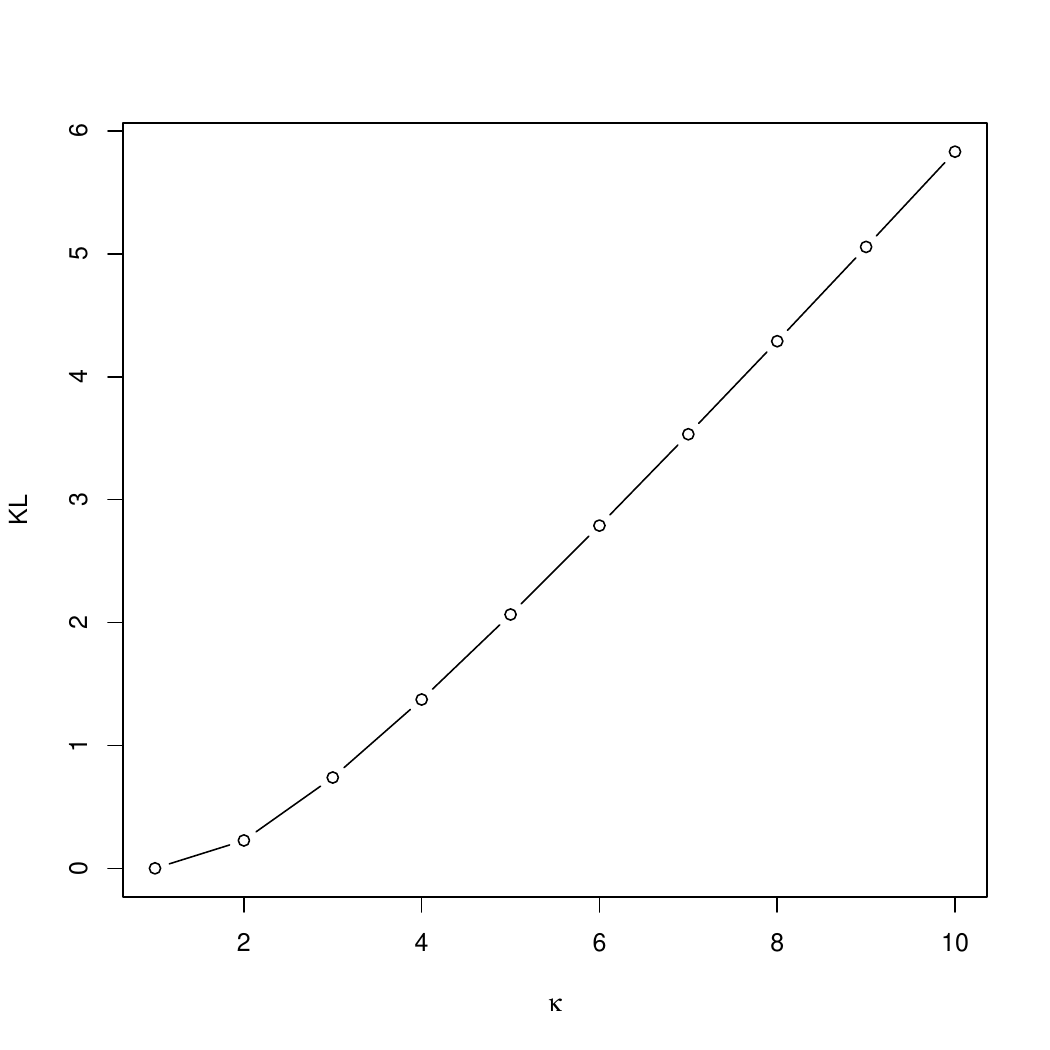}}}
\subfloat[]{%
{\includegraphics[width=0.4\textwidth,height=0.4\textwidth]{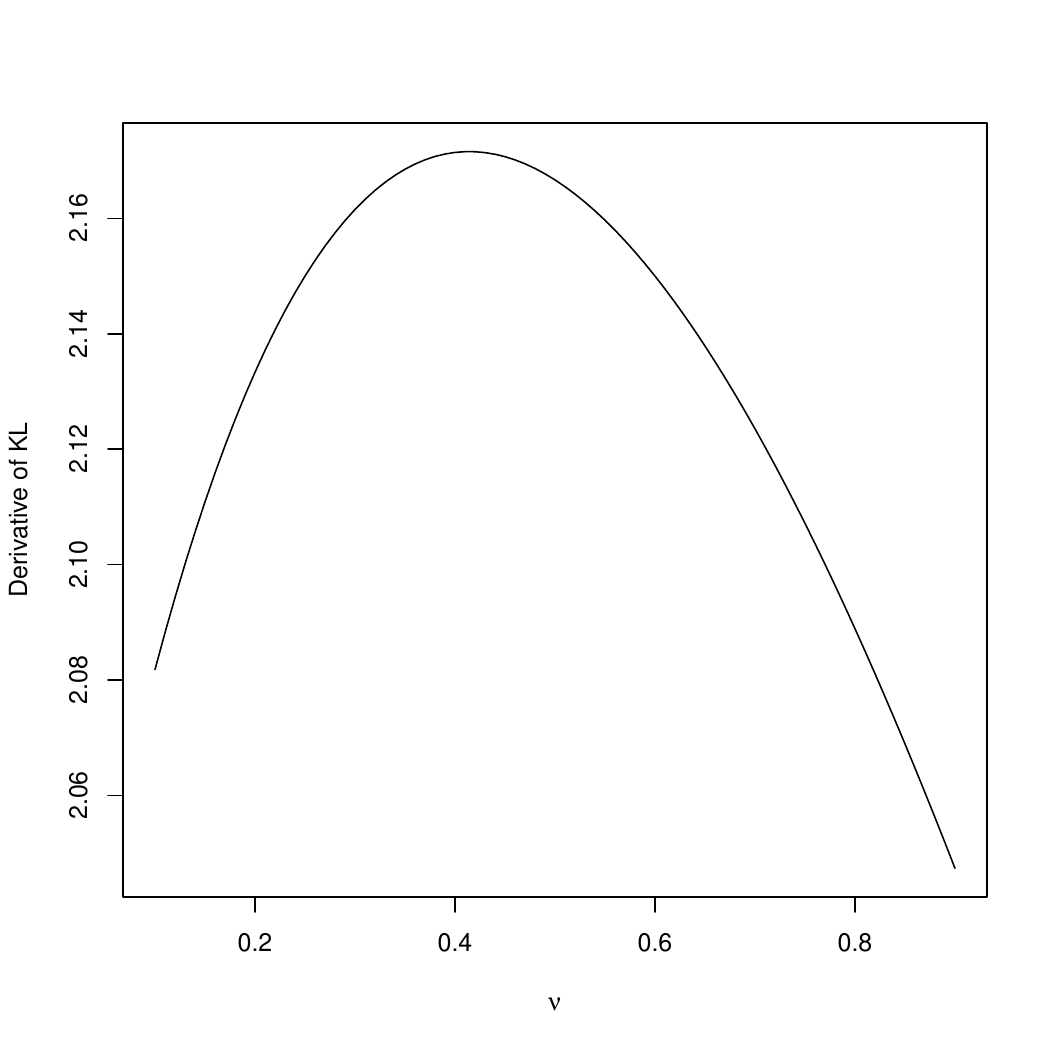}}}
\caption{  (a) The circular variance plot of the symmetric density in Equation (\ref{symm_voncos}).  (b) The plot of the KL divergence $D_{KL}(f_c,h_3)$ between proposed and Cardioid density as in Equation (\ref{symm_kl}) for $\kappa=0,1,\cdots,10$. (c) The plot of the derivative of KL divergence with respect to $\nu \in (0,1)$.} 
\label{circular_variance_kl}
\end{figure}

Let $\theta_1,\cdots, \theta_n$ be the set of i.i.d observations from the probability density function in Equation (\ref{symm_voncos}), with the parameters $\kappa$, and $\nu$. The log-likelihood function is given by 

\begin{eqnarray}
    l&=&\sum_{i=1}^n\kappa \cos \theta_i+\sum_{i=1}^n \log \left[(1+\nu \cos\theta_i)\right]
    -n \log(2\pi)-n\log(I_0(\kappa)+\nu I_1(\kappa))
    \label{symm_log_liklihood}
\end{eqnarray}
Differentiating the log-likelihood function with respect to $\kappa$ and equating it with zero gives

\begin{eqnarray}
    \frac{\partial l}{\partial \kappa}&=&\sum_{i=1}^n\cos\theta_i-n\frac{I_{1}(\kappa)+\frac{\nu}{2}~[I_{0}(\kappa)+I_{2}(\kappa)]}{(I_{0}(\kappa)+\nu ~I_{1}(\kappa))}=0.
\end{eqnarray}
Now, using the fact $\sum_{i=1}^n\cos(\theta_i-\mu_1)=\sum_{i=1}^n\cos\theta_i=\mathcal{R},$ the length of the resultant vector \cite[see][Theorem-1.1, pp. 19]{jammalamadaka2001topics},  the immediate above equation leads to 

\begin{eqnarray}
\frac{A(\kappa)+\nu~[1-\frac{A(\kappa)}{\kappa}]}{(1+\nu ~A(\kappa))}&=&\frac{\mathcal{R}}{n}.
    \label{symm_kappalikeli}
\end{eqnarray}

Differentiating Equation (\ref{symm_log_liklihood}) with respect to $\nu$ and equating with zero provides

\begin{eqnarray}
    \frac{\partial l}{\partial \nu}&=&\sum_{i=1}^n \frac{\cos\theta_i}{[(1+\nu \cos\theta_i)]} -n \frac{I_{1}(\kappa)}{(I_{0}(\kappa)+\nu~I_{1}(\kappa))}=0\nonumber\\
    \sum_{i=1}^n \frac{\cos\theta_i}{[(1+\nu \cos\theta_i)]} &=& \frac{nA(\kappa)}{(1+\nu~A(\kappa))}\\
    \nonumber
    \label{symm_alikeli}
\end{eqnarray}

In this case, the elements of the observed information matrix are
\begin{eqnarray}
   J^{0}_{\kappa\kappa} &=& \frac{n}{(1+\nu A(\kappa))^2} \bigg[ (1+\nu A(\kappa))\left(A'(\kappa) - \nu \left(\frac{\nu A'(\kappa)-A(\kappa)}{\kappa^2} \right) \right) \nonumber \\
    & &\hspace{2cm} -   \nu A'(\kappa) \left (A(\kappa)+\nu~ \left[1-\frac{A(\kappa)}{\kappa}\right]\right)  \bigg]
     \nonumber
\end{eqnarray}

\begin{eqnarray}
    J^{0}_{\nu\nu} &=&- \frac{nA^2(\kappa)}{(1+\nu ~A(\kappa))^2}+\sum_{i=1}^n \frac{\cos^2\theta_i}{[(1+\nu \cos\theta_i)]^2} 
    \nonumber
\end{eqnarray}

\begin{eqnarray}
  J^{0}_{\kappa\nu} &=&n\frac{~ \left[1-\frac{A(\kappa)}{\kappa}\right]-A^2(\kappa) }{(1+\nu A(\kappa))^2},
   \nonumber
\end{eqnarray}

and the elements of expected observation matrix are
\begin{eqnarray}
   \iota^{0}_{\kappa\kappa} &=& \frac{1}{(1+\nu A(\kappa))^2} \bigg[ (1+\nu A(\kappa))\left(A'(\kappa) - \nu \left(\frac{\nu A'(\kappa)-A(\kappa)}{\kappa^2} \right) \right) \nonumber \\
    & &\hspace{2cm} -   \nu A'(\kappa)   \left (A(\kappa)+\nu~ \left[1-\frac{A(\kappa)}{\kappa}\right]\right)  \bigg]
     \nonumber
\end{eqnarray}

\begin{eqnarray}
    \iota^{0}_{\nu\nu}&=& \frac{A^2(\kappa)}{(1+\nu ~A(\kappa))^2}-  \int_{0}^{2\pi}
 \frac{\cos^2\theta}{[(1+\nu \cos\theta)]^2} h_3(\theta) d\theta.
    \nonumber
\end{eqnarray}

\begin{eqnarray}
  \iota^{0}_{\kappa\nu} &=&\frac{ \left[1-\frac{A(\kappa)}{\kappa}\right]-A^2(\kappa) }{(1+\nu  A(\kappa))^2}.
   \nonumber
\end{eqnarray}

Similar to the situation when $\mu \neq 0$, here also one can obtain the numerical solutions for the parameters.

\subsection{ Comparison With the Jones and Pewsey Distribution}
\label{comparison_jones_pew}

Another well-known three-parameter family of symmetric circular
distributions were proposed by \cite{jones2005family}. This
family of distributions has a density
\begin{eqnarray}
    f_{jp}(\theta)&=& \frac{[\cosh{(\kappa\psi)} +\sinh{(\kappa\psi)} \cos(\theta-\mu)]^{1/\psi}}{2\pi P_{1/\psi}(\cosh{(\kappa\psi)})},
    \label{jones_pewsey_dist}
\end{eqnarray}
where $0\leq \theta \leq 2\pi$, $P_{1/\psi}(z)$ is the associated Legendre function of the first with degree $1/\psi$ and order $0$
\cite[see][secs. 8.7 and 8.8]{gradshteyn2014table}, $0\leq \mu <2\pi$, $\kappa > 0$, and $-\infty<\psi<\infty$.  The probability density functions represented by Equation (\ref{symm_voncos}) and Equation (\ref{jones_pewsey_dist}) demonstrate several common features, alongside some differences. Now we will discuss the similarities in the following 
\begin{itemize}
    \item Symmetric about $\theta= 0$.
    \item  Unimodal with mode at $\theta = 0$ and anti mode at     $\theta= \pi$.

    \item  Have the von Mises and Cardioid distributions as
special cases.
\end{itemize}

An advantage of the distribution of \cite{jones2005family} is that it includes a broader range of reconsidered special case densities, including the Wrapped Cauchy and Cartwright’s power-of-cosine distribution. Advantages of the proposed distribution in  Equation (\ref{symm_voncos}) include its attractive properties, such as a closed form of the normalizing constant and trigonometric moments with less complexity in calculating the normalizing constant and trigonometric moments and its ready extension to a three-parameter asymmetric family of distributions.
Along with the von Mises distribution on the surface of the curved torus, we can also introduce some other distributions on it.

\section{Data analysis}
\label{data_analysis}

\subsection{Fitting proposed marginal distribution to the wind direction data }  
Wind direction is a crucial meteorological parameter influencing various weather patterns and climatic conditions. Understanding the variability and trends in wind direction is essential for numerous applications, including agriculture, renewable energy production, urban planning, and air quality management.

This study analyzes wind direction variability in Kolkata (Latitude 22.57, Longitude 88.36), the capital city of West Bengal, India, over a 41-year period during the month of August, from 1982 to 2023.  In total, 1271 daily observations are examined. The primary variable of interest is the 10-meter wind direction (WD10M), obtained from the MERRA-2 reanalysis dataset via NASA's POWER CERES/MERRA2 Native Resolution Daily Data portal (\url{[https://power.larc.nasa.gov/data-access-viewer/}). The full model defined in Equation (\ref{voncos}) is fitted to the data using maximum likelihood estimation (MLE).

Table-\ref{table:wind_data} presents the maximum likelihood estimates (MLE)  of the parameters (standard errors),  the Akaike Information Criterion (AIC) and Bayesian Information Criterion (BIC) values for both the model described in Equation (\ref{voncos}) and the von Mises distribution. Based on the   AIC and BIC criteria, the proposed model described in Equation (\ref{voncos}) stands out to be a preferable one. 
Chi-squared tests assessing the goodness of fit for the distributions, utilizing the binning illustrated in Figure-\ref{wind_data_fig}(a), do not reject the null hypothesis for both the proposed family of distributions outlined in Equation (\ref{voncos}) (p-value=$0.51$) and the von Mises distribution (p-value=$0.32$).

Additionally, a density-based comparison between the proposed distribution Equation (\ref{voncos}) and the von Mises distribution is depicted along with the histogram of the data set in Figure-\ref{wind_data_fig}(a). The fitted densities are based on the maximum likelihood estimates (MLE) of the parameters. Furthermore, a CDF-based comparison can be observed in Figure-\ref{wind_data_fig}(b) between these two models.

Upon both visual inspection and the findings reported in Table-\ref{table:wind_data}, it is apparent that the fitted density from the complete model described in Equation (\ref{voncos}) provides a superior fit compared to the von Mises distribution for this specific dataset.

\begin{table}[h!]
\scalebox{0.85}{
\centering
\begin{tabular}{|l|c c c c c c c |}
\hline
\textbf{MLE of parameters} & $\hat{\kappa}$ & $\hat{\mu}$ & $\hat{\nu}$ & $\log L$ & AIC & BIC & p-value\\
\hline
\hline
\textbf{Proposed model} & $3.47 (0.22) $& 3.09 (0.02) & 0.66 (0.07) & $-1444.03$  & $2894.23$ & $2909.50$ & 0.51 \\

\hline
\textbf{von Mises } & $2.40 (0.09)$  & 3.07 (0.02)  &  & $ -1448.41$ & $2901.81$   & $2911.81$  & $0.32$ \\
\hline
\end{tabular}}
\vspace{.1cm}
\caption{Maximum likelihood estimates (MLE) of the parameters (standard errors), the maximized log-likelihood, Akaike information criterion (AIC), Bayes information criterion(BIC), and p-value of the Chi-squared test statistic for the  proposed in Equation (\ref{voncos}), and one its submodels, the von Mises distribution.}
\label{table:wind_data}
\end{table}

\begin{figure}[h!]
\centering
\subfloat[]{%
{\includegraphics[width=0.55\textwidth,height=0.55\textwidth]{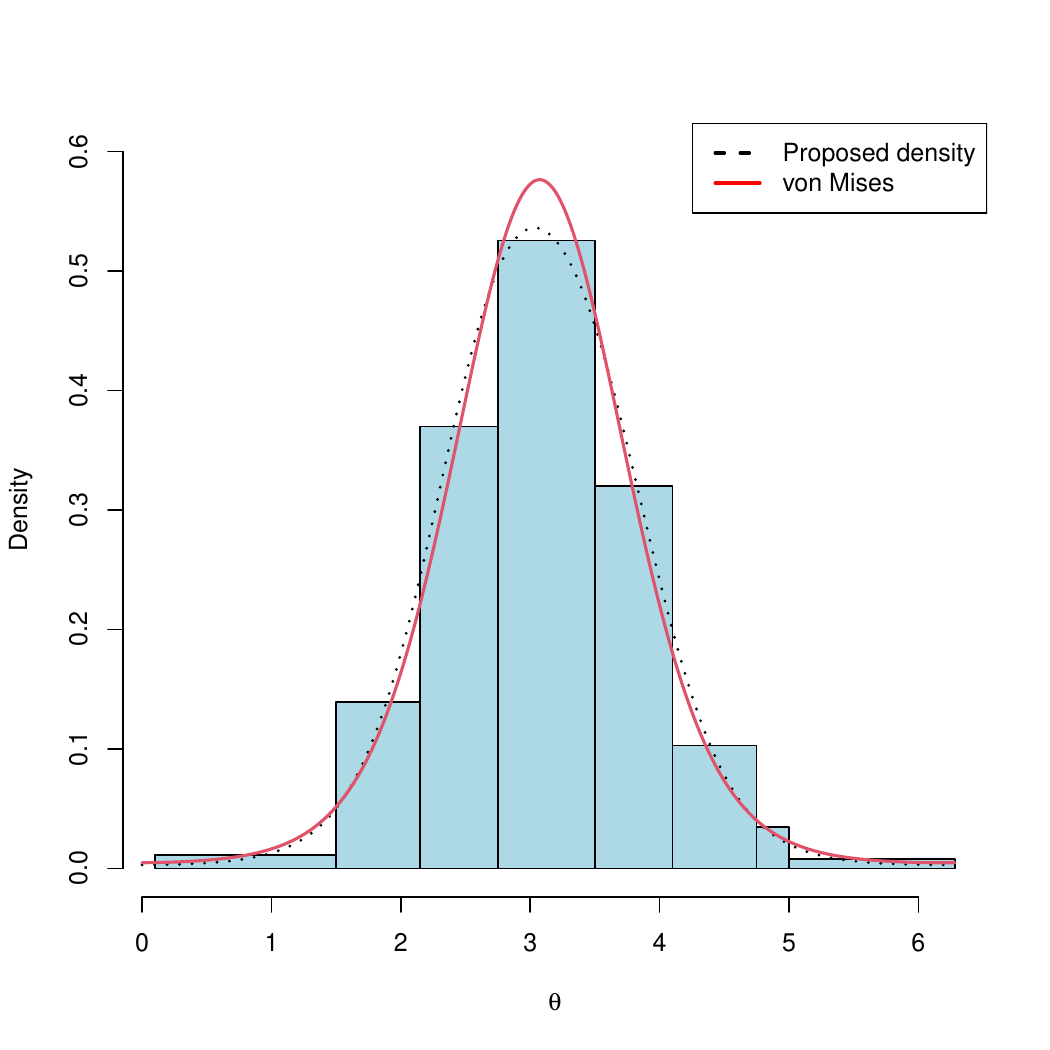}}}
\subfloat[ ]{%
{\includegraphics[width=0.52\textwidth,height=0.52\textwidth]{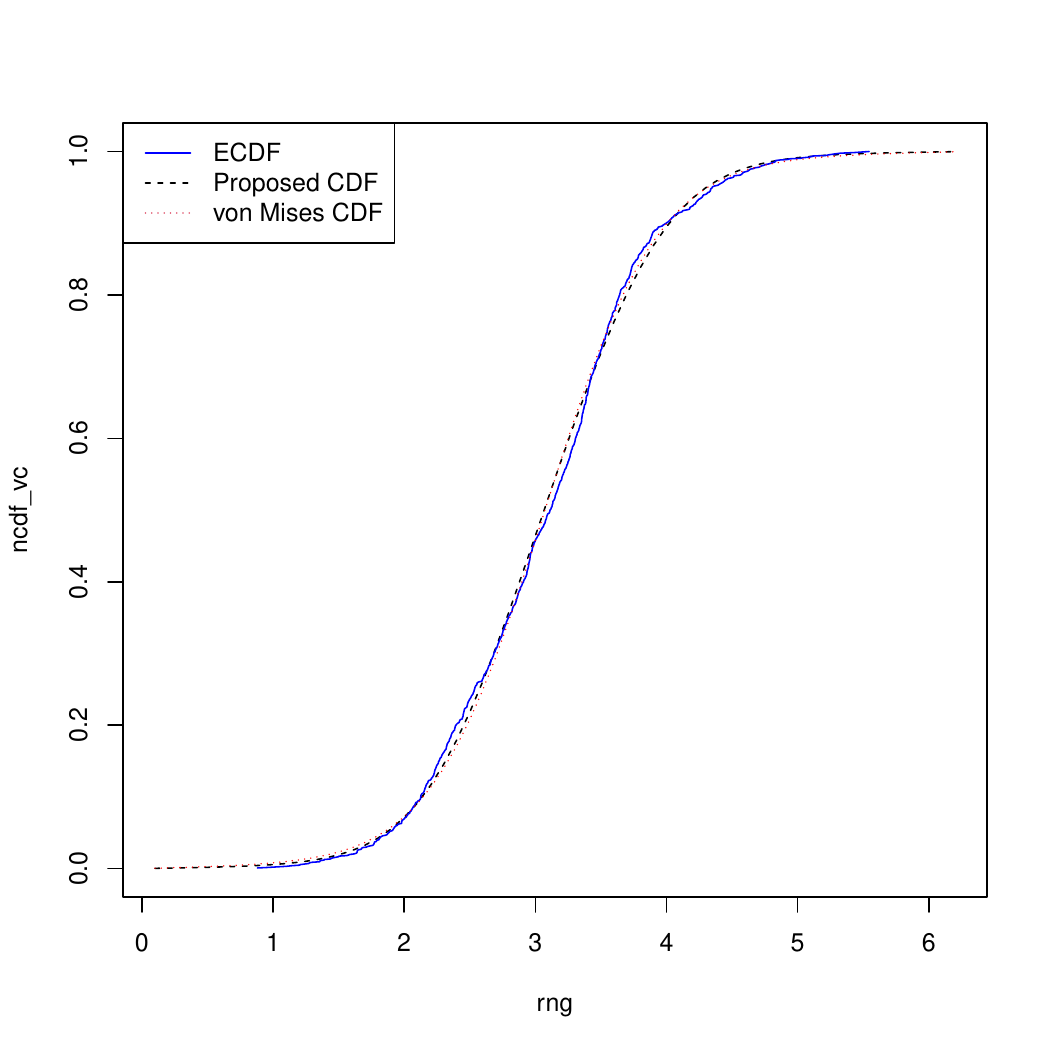}}}\hspace{5pt}
\caption{The figure depicts the wind direction variability at Kolkata, India, over 41 years, specifically for the month of August from the year 1982 to 2023. 
(a) The histogram of the data of wind directions along with the estimated density obtained through the maximum likelihood estimate (MLE) of the parameters of the general family of distributions according to Equation (\ref{voncos}). Additionally, the two-parameter sub-model, represented by the von Mises density, is displayed as solid lines.
(b) The plot of the fitted cumulative distribution functions (CDFs) of the proposed distribution in Equation (\ref{voncos}) and von Mises distribution, along with the empirical cumulative distribution function (ECDF) of the data.} 
\label{wind_data_fig}
\end{figure}

\section{Conclusion}
\label{conclusion}
This study presents a computationally efficient approach for large-scale sampling from circular distributions. The same is extended to the framework of toroidal distributions on the surface of a curved torus. The key to the development lies in enhancing the acceptance-rejection sampling method based on the upper Riemann sum. This method addresses a long-standing issue in rejection sampling—the selection of a suitable proposal density by constructing a piecewise constant envelope that naturally dominates the target density. The resulting proposal distribution, being both tractable and theoretically justifiable, significantly reduces rejection rates and improves computational performance. Compared to the classical sampling algorithm by \cite{best1979efficient} for the von Mises distribution, the proposed method yields higher acceptance rates with lower computational cost, making it particularly valuable for large-scale Monte Carlo methods.

Beyond sampling, this work contributes to the growing body of literature on distributions defined over manifolds by exploring the geometry-driven construction of probability models on the curved torus, $\mathbb{S}^1 \times \mathbb{S}^1$. Motivated by the work of \cite{diaconis2013sampling}, which emphasized the role of intrinsic geometry in  deriving maximum entropy distributions (e.g., von Mises on the circle, Fisher on the sphere), we construct a toroidal analogue on the surface of the curved torus.

In summary, this paper bridges two key areas: efficient sampling from circular and toroidal distributions and the construction of intrinsic maximum entropy distributions on the curved torus. The proposed sampling algorithm, coupled with the geometry-driven toroidal model, sets the stage for robust inferential methods in applications involving bivariate angular data. Future work may focus on extending this framework to dependent models on the torus and incorporating covariate information in regression settings.
\section*{Competing interests}
No competing interest is declared.



\section*{Acknowledgement} The first author acknowledges and appreciates the financial assistance provided in the form of a junior/senior research fellowship by the Ministry of Human Resource and Development (MHRD) and IIT Kharagpur, India.  Both the authors would like to thank the anonymous referee for sharing the valuable comments on the article, which have helped us to enhance the manuscript a lot.


\bibliographystyle{agsm}
 \bibliography{buddha_bib}

 \newpage

 \section{Appendix}
 \label{appendix}

\subsection{Intrinsic geometry of torus} \label{geometry_torus}
The parametric equation of $2$-dimensional curved torus  is the Lipschitz image \cite[see][]{diaconis2013sampling} of the set $\{ 
 (\phi,\theta):0<\phi,\theta<2\pi\}\subset \mathbb{R}^2.$ Clearly, the function $f(\phi,\theta)=\{  (R+r\cos{\theta})\cos{\phi}, (R+r\cos{\theta})\sin{\phi}, r\sin{\theta} \}$ is a differentiable function from $\mathbb{R}^2$ to $\mathbb{R}^3$. Now, the partial derivatives of $f$ with respect to $\phi$, and $\theta$ are 
$$ \dfrac{\partial f}{\partial \phi}=\{ -(R+r\cos{\theta})\sin{\phi}, (R+r\cos{\theta})\cos{\phi}, 0 \}, $$ and
$$\dfrac{\partial f}{\partial \theta}=\{  -r\sin{\theta} \cos{\phi}, -r\sin{\theta}\sin{\phi}, r\cos{\theta}  \},$$ respectively. Hence, the derivative matrix is

$$Df(\phi,\theta)=\begin{bmatrix}
 -(R+r\cos{\theta})\sin{\phi} &  -r\sin{\theta} \cos{\phi} \\
 (R+r\cos{\theta})\cos{\phi} & -r\sin{\theta}\sin{\phi}\\
  0& r\cos{\theta}
\end{bmatrix}.$$ 

Therefore, the square of the Jacobian can be calculated as

\begin{eqnarray}
       J_{2}^{2}f(\phi,\theta)&=& \text{det} \left[ D^Tf(\phi,\theta) \cdot Df(\phi,\theta)\right]\nonumber\\
       &=& \text{det} \begin{bmatrix}
 (R+r\cos{\theta})^{2} & 0 \\
 0 & r^{2}
\end{bmatrix} \nonumber\\
&=&r^{2}(R+r\cos{\theta})^{2}\nonumber\\
\label{jacobian}
\end{eqnarray}
Using the above expression, we get the area element as 
\begin{equation}
    dA=r(R+r\cos{\theta}) d\phi d\theta,
    \label{area_element}
\end{equation}
 which is the square root of the determinant of the product of transpose of the derivative matrix with itself.

\subsection{The Normalizing Constant} \label{appendix har}

 We will determine the normalizing constant C for the von Mises distribution on a torus' surface. We will utilize the subsequent two Bessel function identities to achieve this objective.

\begin{equation}
    \frac{1}{2\pi}\int_{0}^{2\pi} e^{\kappa\cos{\theta}} \cos{p\theta}~d\theta=I_p(\kappa),
    \label{i0k}
\end{equation}
and
\begin{equation}
   \frac{1}{2\pi}\int_{0}^{2\pi} e^{\kappa\cos{\theta}} \sin{n\theta}~d\theta=0.
   \label{ink}
\end{equation}

Now, the constant $C$ can be written as
\begin{eqnarray}
    C&=& \displaystyle \int_{0}^{2\pi} \left[e^{\kappa\cos(\theta-\mu)}\left(1+\nu\cos\theta \right)\right]~d\theta \nonumber\\
     &=&\displaystyle \int_{0}^{2\pi} e^{\kappa\cos(\theta-\mu)}~d\theta + \nu \int_{0}^{2\pi} e^{\kappa\cos(\theta-\mu)} \cos\theta ~d\theta \nonumber\\
     &=& C_1+C_2,
     \label{c}
\end{eqnarray}
where using Equation (\ref{i0k}) we get
\begin{equation}
    C_1=\displaystyle \int_{0}^{2\pi} e^{\kappa\cos(\theta-\mu)}~d\theta=2\pi I_0(\kappa),
    \label{c1}
\end{equation}
 and 

$$C_2=\nu\int_{0}^{2\pi} e^{\kappa\cos{(\theta-\mu)}}\cos\theta ~d\theta .$$

Now for $C_2$, let $\theta-\mu=t$ $\implies~ d\theta=dt, \mbox{~and ~}\theta=\mu+t.$ Therefore, $C_2$ becomes
\begin{eqnarray}
    C_2&=& \nu\int_{-\mu}^{2\pi-\mu} e^{\kappa\cos t}\cos {(\mu+t)} ~dt  \nonumber\\
    &=&\nu\int_{-\mu}^{2\pi-\mu} e^{\kappa\cos t}\left( \cos t \cos \mu -\sin t \sin \mu \right)~dt \nonumber\\
    &=&  \nu \cos \mu \int_{-\mu}^{2\pi-\mu} e^{\kappa\cos t} \cos t ~dt +  \nu \sin \mu \int_{-\mu}^{2\pi-\mu} e^{\kappa\cos t} \sin t ~dt \nonumber
\end{eqnarray}

Now, using the identities from Equation (\ref{i0k}), and Equation (\ref{ink}) we can write
\begin{equation}
    C_2=\nu \cos \mu ~2\pi I_1(\kappa).
    \label{c2}
\end{equation}

Substituting the values from the Equation (\ref{c1}), and Equation (\ref{c2}) in the Equation (\ref{c}) we get
 \begin{equation*}
     C=2\pi \left[I_0(\kappa)+ \nu \cos \mu ~I_1(\kappa) \right].
     \label{ccom}
 \end{equation*}

\subsection{Proof of Theorem-\ref{max_entropy_thm}} \label{max_entropy_proof}

\begin{proof}

See the Appendix for the proof.
    Let $\Psi$ be any arbitrary circular random variable having the probability density function $q(\psi)$ with unit circle being the support which satisfies the above constraints. Now, let us define the entropy and cross-entropy as follows:
\begin{equation}
    H_q(q(\theta))=-\int_{0}^{2\pi} q(\theta)\log(  q(\theta)) ~d\theta
    \label{entropy}
\end{equation}

\begin{equation}
    H_q(h_2(\theta))=-\int_{0}^{2\pi} q(\theta)\log(h_2(\theta)) ~d\theta
    \label{cross entropy}
\end{equation}

 The KL divergence between $f$ and $q$ is given by 

\begin{eqnarray}
     D_{KL}(q||h_2(\theta))&=&\int_{0}^{2\pi} q(\theta)\log \left[\frac{q(\theta)}{h_2(\theta)}\right] ~d\theta \nonumber\\
     &=&H_q(h_2(\theta))-H_q(q(\theta))\nonumber\\ 
     H_q(q(\theta))&=&H_q(h_2(\theta))- D_{KL}(q||h_2(\theta))
     \label{KLdiv}
\end{eqnarray}

Consider the following from the Equation (\ref{voncos}) and Equation (\ref{jacobian})

\begin{eqnarray}
   \log(h_2(\theta))&=&\kappa\cos(\theta-\mu) + \log \left( 1+\nu\cos\theta \right)- \log\left[2\pi(I_{0}(\kappa)+\nu \cos \mu~I_{1}(\kappa)) \right]\nonumber\\
    &=&\kappa\cos(\theta-\mu) + \log(J_2(f(\phi,\theta)))-\log(rR)- \log\left[2\pi \left(I_{0}(\kappa)+\nu \cos \mu~I_{1}(\kappa)\right) \right]  \nonumber\\
      &=&\kappa\cos(\theta-\mu) + \log(J_2(f(\phi,\theta)))- \log\left[2\pi ~rR~\left(I_{0} (\kappa)+\nu \cos \mu~I_{1}(\kappa) \right) \right]  \nonumber\\
      &=&\kappa\cos(\theta-\mu) + \log(J_2(f(\phi,\theta)))+a_4, \nonumber\\
\end{eqnarray}
where $a_4=- \log\left[2\pi ~rR~(I_{0}(\kappa)+\nu \cos \mu~I_{1}(\kappa)) \right]$ is a fixed constant. This leads us to the following

\begin{eqnarray}
     H_q(h_2(\theta))&=&-\int_{0}^{2\pi} \left[ \kappa\cos(\theta-\mu) + \log(J_2(f(\phi,\theta)))+a_4 \right] q(\theta) ~d\theta  \nonumber\\
     &=& \kappa~\cos \mu ~E(  \cos\theta) + \kappa~\sin \mu ~E(  \sin\theta) +E(\log(J_2(f(\phi,\theta))))+a_4 \nonumber\\
     &=& \kappa~\cos \mu ~E(  \cos\theta) + \kappa~\sin \mu ~E(  \sin\theta) +E(\log(\nu(1+\nu \cos \theta))+a_4 \nonumber\\
     &\leq& \kappa~\cos \mu ~E(  \cos\theta) + \kappa~\sin \mu ~E(  \sin\theta) +E(\log(\nu(1+\nu))+a_4 \nonumber\\
     &=& \kappa~\cos \mu ~E(  \cos\theta) + \kappa~\sin \mu ~E(  \sin\theta) +\log(\nu(1+\nu))+a_4 \nonumber\\
      &=& \kappa a_1 ~\cos \mu + \kappa a_2 ~\sin \mu  +a_3+a_4 \nonumber\\
    \implies H_q(h_2(\theta))  &\leq& H_{h_2(\theta)}(h_2(\theta)).
\end{eqnarray}

Using this in the   Equation (\ref{KLdiv}) we have 

\begin{eqnarray}
    H_q(q(\theta))&=&H_q(h_2(\theta))- D_{KL}(q||h_2(\theta)) \nonumber\\
    H_q(q(\theta))&\leq&H_{h_2(\theta)}(h_2(\theta))- D_{KL}(q||h_2(\theta)) \nonumber\\
     H_q(q(\theta))&\leq&H_{h_2(\theta)}(h_2(\theta)).\nonumber\\
\end{eqnarray}
Hence, the theorem follows.
\end{proof}

\subsection{Proof of Theorem-\ref{ch.function}} \label{ch_fun_proof}

\begin{proof}
    Consider the following integral 

\begin{eqnarray}
    J&=& \int_{0}^{2\pi} e^{\kappa\cos(\theta-\mu)}\left(1+\nu\cos\theta \right)e^{ip\theta} ~d\theta\nonumber\\
    &=&\int_{0}^{2\pi} e^{\kappa\cos(\theta-\mu)} e^{ip\theta} ~d\theta +\nu\int_{0}^{2\pi} \left(e^{\kappa\cos(\theta-\mu)}\cos\theta \right)e^{ip\theta} ~d\theta\nonumber\\
    &=&J_1+\nu J_2,
    \label{j1+ratioj2}
\end{eqnarray}
where \begin{eqnarray}
    J_1&=&\int_{0}^{2\pi} e^{\kappa\cos(\theta-\mu)} e^{ip\theta} ~d\theta  \label{ch.vonpart}\\
    J_2&=&\int_{0}^{2\pi} \left(e^{\kappa\cos(\theta-\mu)}\cos\theta \right)e^{ip\theta} ~d\theta
    \label{ch.cospart}
\end{eqnarray}

Consider the Equation (\ref{ch.vonpart}) and using the identities in Equation (\ref{i0k}), Equation (\ref{ink}) we get
\begin{eqnarray}
    J_1&=&\int_{0}^{2\pi} e^{\kappa\cos(\theta-\mu)} \left[ \cos p\theta+i \sin p\theta \right] ~d\theta   \nonumber\\
    &=& \int_{0}^{2\pi} e^{\kappa\cos t} e^{i(t+\mu)p} ~d\theta=e^{ip\mu} \int_{0}^{2\pi} e^{\kappa\cos t}  \left[ \cos t+i \sin t \right] ~d\theta \nonumber\\ 
    &=& 2\pi ~e^{ip\mu}~ I_{p}(\kappa)
    \label{J1comp}
\end{eqnarray}

Consider the Equation (\ref{ch.cospart}) and using the identities in Equation (\ref{i0k}), Equation (\ref{ink}) we get

\begin{eqnarray}
    J_2&=&\int_{0}^{2\pi} e^{\kappa\cos(\theta-\mu)} \cos \theta \left[ \cos p\theta+i \sin p\theta \right] ~d\theta   \nonumber\\
    &=&\int_{0}^{2\pi} e^{\kappa\cos(\theta-\mu)} \cos \theta  \cos p\theta ~d\theta+i \int_{0}^{2\pi} e^{\kappa\cos(\theta-\mu)} \cos \theta  \sin p\theta~d\theta \nonumber\\ 
    &=&J_{21}+i ~J_{22},
    \label{J2}
\end{eqnarray}

where, 

\begin{eqnarray}
    J_{21}&=&\int_{0}^{2\pi} e^{\kappa\cos(\theta-\mu)} \cos \theta  \cos p\theta ~d\theta \label{J21} \\
    J_{22}&=& \int_{0}^{2\pi} e^{\kappa\cos(\theta-\mu)} \cos \theta  \sin p\theta~d\theta 
    \label{J22}
\end{eqnarray}
Consider the following from Equation (\ref{J21})

\begin{eqnarray}
    J_{21}&=&\int_{0}^{2\pi} e^{\kappa\cos(\theta-\mu)} \cos \theta  \cos p\theta ~d\theta  \nonumber \\
  &=& \frac{1}{2}\int_{0}^{2\pi} e^{\kappa\cos(\theta-\mu)}  \left[ \cos(p+1) \theta +\cos(p-1) \theta \right] ~d\theta \nonumber \\ 
&=& \frac{1}{2}\int_{0}^{2\pi} e^{\kappa\cos(\theta-\mu)}   \cos(p+1) \theta ~d\theta  + \frac{1}{2}\int_{0}^{2\pi} e^{\kappa\cos(\theta-\mu)} \cos(p-1) \theta ~d\theta \nonumber \\
&=& J_{211}+J_{212}, 
\label{J_{211}+J_{212}}
\end{eqnarray}

where

\begin{eqnarray}
    J_{211}&=&\frac{1}{2}\int_{0}^{2\pi} e^{\kappa\cos(\theta-\mu)}   \cos(p+1) \theta ~d\theta \label{j211} \\
J_{212}&=&   \frac{1}{2}\int_{0}^{2\pi} e^{\kappa\cos(\theta-\mu)} \cos(p-1) \theta ~d\theta. \label{j212}
\end{eqnarray}

Consider the following Equation (\ref{j211}) and the identity in Equation (\ref{ink}) we get

\begin{eqnarray}
    J_{211}&=&\frac{1}{2}\int_{0}^{2\pi} e^{\kappa\cos(\theta-\mu)}   \cos(p+1) \theta ~d\theta \nonumber\\
    &=&\frac{1}{2}\int_{-\mu}^{2\pi-\mu} e^{\kappa\cos t}   \cos(p+1) (\mu+t) ~d\theta \nonumber\\
    &=&\frac{1}{2}\int_{-\mu}^{2\pi-\mu} e^{\kappa\cos t} \left[   \cos(p+1) \mu ~\cos(p+1) t -\sin(p+1) \mu ~\sin(p+1) t \right]~d\theta \nonumber\\
    &=&\frac{1}{2}\int_{-\mu}^{2\pi-\mu} e^{\kappa\cos t} \left[   \cos(p+1) \mu ~\cos(p+1) t \right]~d\theta \nonumber\\
     &=&\frac{\cos(p+1) \mu}{2}\int_{0}^{2\pi} e^{\kappa\cos t} \left[   \cos(p+1) t \right]~d\theta \nonumber\\
    J_{211} &=&\frac{\cos(p+1) \mu}{2}~ 2\pi~ I_{p+1}(\kappa)
\end{eqnarray}
Similarly, we can show that 

\begin{eqnarray}
       J_{212} &=&\frac{\cos(p-1) \mu}{2}~ 2\pi~ I_{p-1}(\kappa)
\end{eqnarray}

Using the values of $J_{211} \& J_{212}$ in Equation (\ref{J_{211}+J_{212}}) we get

\begin{equation}
    J_{21}=\pi\left[ I_{p-1}(\kappa)~\cos(p-1) \mu+ I_{p+1}(\kappa)~\cos(p+1) \mu \right]
    \label{j21comp}
\end{equation}

Consider the following from Equation (\ref{J22})

\begin{eqnarray}
    J_{22}&=&\int_{0}^{2\pi} e^{\kappa\cos(\theta-\mu)} \cos \theta  \sin p\theta ~d\theta  \nonumber \\
  &=& \frac{1}{2}\int_{0}^{2\pi} e^{\kappa\cos(\theta-\mu)}  \left[ \sin(p+1) \theta +\sin(p-1) \theta \right] ~d\theta \nonumber \\ 
&=& \frac{1}{2}\int_{0}^{2\pi} e^{\kappa\cos(\theta-\mu)}   \sin(p+1) \theta ~d\theta  + \frac{1}{2}\int_{0}^{2\pi} e^{\kappa\cos(\theta-\mu)} \sin(p-1) \theta ~d\theta \nonumber \\
&=& J_{221}+J_{222}, 
\label{J_{221}+J_{222}}
\end{eqnarray}

where

\begin{eqnarray}
    J_{221}&=&\frac{1}{2}\int_{0}^{2\pi} e^{\kappa\cos(\theta-\mu)}   \sin(p+1) \theta ~d\theta \label{j221} \\
J_{222}&=&   \frac{1}{2}\int_{0}^{2\pi} e^{\kappa\cos(\theta-\mu)} \cos(p-1) \theta ~d\theta. \label{j222}
\end{eqnarray}

Consider the following Equation (\ref{j221}) and the identity in Equation (\ref{ink}) we get

\begin{eqnarray}
    J_{221}&=&\frac{1}{2}\int_{0}^{2\pi} e^{\kappa\cos(\theta-\mu)}   \sin(p+1) \theta ~d\theta \nonumber\\
    &=&\frac{1}{2}\int_{-\mu}^{2\pi-\mu} e^{\kappa\cos t}   \sin(p+1) (\mu+t) ~dt \nonumber\\
    &=&\frac{1}{2}\int_{-\mu}^{2\pi-\mu} e^{\kappa\cos t} \left[   \sin(p+1) \mu ~\cos(p+1) t +\cos(p+1) \mu ~\sin(p+1) t \right]~dt \nonumber\\
    &=&\frac{1}{2}\int_{-\mu}^{2\pi-\mu} e^{\kappa\cos t} \left[   \sin(p+1) \mu ~\cos(p+1) t \right]~dt \nonumber\\
     &=&\frac{\sin(p+1) \mu}{2}\int_{0}^{2\pi} e^{\kappa\cos t} \left[   \cos(p+1) t \right]~dt \nonumber\\
    J_{221} &=&\frac{\sin(p+1) \mu}{2}~ 2\pi~ I_{p+1}(\kappa)
\end{eqnarray}
Similarly, we can show that 

\begin{eqnarray}
       J_{222} &=&\frac{\sin(p-1) \mu}{2}~ 2\pi~ I_{p-1}(\kappa)
\end{eqnarray}

Using the values of $J_{221} \& J_{222}$ in Equation (\ref{J_{221}+J_{222}}) we get

\begin{equation}
    J_{22}=\pi\left[ I_{p-1}(\kappa)~\sin(p-1) \mu+ I_{p+1}(\kappa)~\sin(p+1) \mu \right]
    \label{j22comp}
\end{equation}

Using the values of $J_{21}\& J_{22}$ in Equation (\ref{J2}) we get

\begin{eqnarray}
       J_{2}&=& \pi\left[ I_{p-1}(\kappa)~\cos(p-1) \mu+ I_{p+1}(\kappa)~\cos(p+1) \mu \right] \nonumber\\
      &+& i~\pi\left[ I_{p-1}(\kappa)~\sin(p-1) \mu+ I_{p+1}(\kappa)~\sin(p+1) \mu \right]\nonumber\\
      &=&\pi I_{p-1}(\kappa)e^{i(p-1)\mu}+\pi  I_{p+1}(\kappa)e^{i(p+1)\mu}
      \label{J2comp}
\end{eqnarray}

Putting the values of $J_1 \& J_2$ in Equation (\ref{j1+ratioj2}) we get

\begin{equation}
    J=\pi \nu I_{p-1}(\kappa)e^{i(p-1)\mu}+2\pi~I_{p}(\kappa)e^{ip\mu}+\pi \nu   I_{p+1}(\kappa)e^{i(p+1)\mu}.
\end{equation}
Therefore the $p^{th}$ trigonometric moments  for $p=0,\pm 1, \pm2, \ldots$ are given by 

\begin{eqnarray}
    \Phi_{p}&=& \frac{J}{2\pi\left[I_{0}(\kappa)+\nu \cos \mu~I_{1}(\kappa)\right]}  \nonumber\\
    &=&\frac{\pi \nu I_{p-1}(\kappa)e^{i(p-1)\mu}+2\pi~I_{p}(\kappa)e^{ip\mu}+\pi \nu   I_{p+1}(\kappa)e^{i(p+1)\mu}}{2\pi\left[I_{0}(\kappa)+\nu \cos \mu~I_{1}(\kappa)\right]}  \nonumber\\
    \Phi_{p}&=&\frac{\nu I_{p-1}(\kappa)e^{i(p-1)\mu}+2~I_{p}(\kappa)e^{ip\mu}+\nu   I_{p+1}(\kappa)e^{i(p+1)\mu}}{2\left[I_{0}(\kappa)+\nu \cos \mu~I_{1}(\kappa)\right]} 
\end{eqnarray}
Hence, the theorem follows.
\end{proof}

\subsection{Proof of Corollary-\ref{symm_con}}
\label{symm_con_proof}

\begin{proof}
Clearly the density in Equation (\ref{voncos}) is symmetric if $\mu=0$, $\kappa=0$ or $\nu \xrightarrow{} 0$. We consider the necessary condition for symmetry. Let $h_2$ be the density in Equation (\ref{voncos}) and assume that $\mu \neq 0$, $\kappa \neq 0$ or $\nu \xrightarrow{} 0$. The density function is symmetric if there exists a constant $v\in [0,2\pi)$ such that $h_2(v+\theta)=h_2(v-\theta)$ for all $\theta \in [0,2\pi)$.
Consider the following

\begin{eqnarray}
    h_2(v+\theta)-h_2(v-\theta) &=& \frac{1}{2\pi(I_{0}(\kappa)+\nu \cos \mu~I_{1}(\kappa))}\left[ e^{\kappa\cos{(v+\theta-\mu)}}\left(1+\nu\cos{(v+\theta)} \right) \right. \nonumber \\
    && \hspace{3cm} -\left.  e^{\kappa\cos(v-\theta-\mu)}\left(1+\nu\cos{(v-\theta)}\right) \right] \nonumber \\
       &=&\frac{e^{\kappa\cos{(v+\theta-\mu)}}\left(1+\nu\cos{(v+\theta)} \right)}{2\pi(I_{0}(\kappa)+\nu \cos \mu~I_{1}(\kappa))}\left[ 1- Q(\theta)  \right], \nonumber\\
\end{eqnarray}
 where 

$$
Q(\theta) = \frac{\left[1 + \nu\cos{(v-\theta)}\right]}{\left[1 + \nu\cos{(v+\theta)}\right]} \times \exp\left\{\kappa\left[\cos{(v+\theta-\mu)} - \cos{(v+\theta+\mu)}\right]\right\}
$$
If $h_2(\theta)$ is symmetric then $Q(\theta)=1$ for all $\theta \in [0,2\pi)$. Our aim is to show that 
$\log [Q(\theta)] \neq 0$ for some $\theta \in [0,2\pi)$.

Now $$\log[Q(\theta)]=\log \left[\frac{1 + \nu\cos{(v-\theta)}}{1 + \nu\cos{(v+\theta)}}\right] + \kappa\left[\cos{(v+\theta-\mu)} - \cos{(v+\theta+\mu)}\right]$$

As by the assumption $\nu\xrightarrow{}0$, the first term  implies that $v=0 \mbox{~or~} \pi.$ Since, $\kappa \neq 0$ this leads to 

$N(\theta) \equiv  \left[\cos{(v+\theta-\mu)} - \cos{(v+\theta+\mu)}\right]$ should be $0$ for any $\theta \in [0,2\pi)$. When $v=0$ then $N(\theta)$ can be written as $N(\theta) =  \left[\cos{(\theta-\mu)} - \cos{(\theta+\mu)}\right]$, and since we have assumed that $\mu\neq0$, $N(\theta) \neq 0$ for some $\theta \in [0,2\pi)$. Similarly, it can be shown that  $N(\theta) \neq 0$ for some $\theta \in [0,2\pi)$ for $v=\pi.$ Hence, the probability density function $h_2(\theta)$ is not symmetric.
\end{proof}

\subsection{Calculating Kullback-Leibler (KL) divergence}
\label{kl_appendix}
\begin{eqnarray}
    D_{KL}(f_c,h_2)&=&\int_{0}^{2\pi}\left[f_c \log \left(\frac{ f_c(\theta) }{h_2(\theta)}\right)  \right
    ]~d\theta \nonumber\\
    &=& \int_{0}^{2\pi}\left[ \frac{1}{2\pi}\left(1+\nu\cos\theta \right) \log \left(\frac{\frac{1}{2\pi}\left(1+\nu\cos\theta \right)}{\frac{e^{\kappa{\cos(\theta-\mu)}}\left(1+\nu\cos\theta \right)}{2\pi(I_{0}(\kappa)+\nu\cos \mu~I_{1}(\kappa))}}\right)  \right
    ]~d\theta \nonumber\\
     &=& \int_{0}^{2\pi}\left[ \frac{1}{2\pi}\left(1+\nu\cos\theta \right) \left[ \log((I_{0}(\kappa)+\nu\cos \mu~I_{1}(\kappa)))-\kappa \cos {(\theta-\mu)}  \right]   \right]~d\theta \nonumber\\
      &=& \log((I_{0}(\kappa)+\nu\cos \mu~I_{1}(\kappa)))-\frac{\kappa}{2\pi} \int_{0}^{2\pi} \left(1+\nu\cos\theta \right) \cos {(\theta-\mu)} ~d\theta \nonumber\\
       &=& \log((I_{0}(\kappa)+\nu\cos \mu~I_{1}(\kappa)))-\frac{\nu\kappa}{2\pi} \int_{0}^{2\pi} \cos\theta \cos {(\theta-\mu)} ~d\theta \nonumber\\
        &=& \log((I_{0}(\kappa)+\nu\cos \mu~I_{1}(\kappa)))-\frac{\nu\kappa}{2\pi} \int_{0}^{2\pi} \cos\theta [\cos \theta \cos \mu - \sin \theta \sin \mu ] ~d\theta \nonumber\\
        &=& \log((I_{0}(\kappa)+\nu\cos \mu~I_{1}(\kappa)))-\frac{\nu\kappa\cos \mu}{4\pi} \int_{0}^{2\pi} [1+\cos 2\theta]~d\theta+\frac{\nu\kappa\sin \mu}{4\pi} \int_{0}^{2\pi} \sin 2\theta  ~d\theta \nonumber\\
        &=& \log((I_{0}(\kappa)+\nu\cos \mu~I_{1}(\kappa)))-\frac{\nu\kappa~\cos \mu}{2} \nonumber\\
        \label{divergence}
\end{eqnarray}

\subsection{} \label{appendix acceptence}
 Consider the cumulative distribution function (CDF) of $X$ for the $i^{th}$ cell as:

 \begingroup
\allowdisplaybreaks
\begin{align}
P(X\leq x)&=P(Y_i\leq x \,| \, Y_i  \hspace{.2cm}\text{accepted}) \nonumber \\
&=\dfrac{P\left(Y_i\leq x, \hspace{.2cm} U<\dfrac{f(Y_i)/P(A_i)}{M_i(\frac{1}
{B})}\right) }{P\left(U<\dfrac{f(Y_i)/P(A_i)}{M_i(\frac{1}
{B})}\right) }
  \label{rejection sampling1i}  
\end{align}%
\endgroup
Now considering the numerator
\begin{equation}
    \begin{aligned}
P\left(Y_i\leq x, \hspace{.2cm} U<\dfrac{f(Y_i)/P(A_i)}{M_i(\frac{1}
{B})}\right) &=\int P\left(Y_i\leq x, \hspace{.2cm} U<\dfrac{f(Y_i)/P(A_i)}{M_i(\frac{1}
{B})} \middle | Y_i=y \right)\left( \frac{1}{B}\right) dy  \\
&=\int \textbf{I}_{(y\leq x)} P\left( U<\dfrac{f(y)/P(A_i)}{M_i(\frac{1}
{B})}\right) \left( \frac{1}{B}\right) dy.\\
\end{aligned}
\label{rejection sampling1i1}
\end{equation}

Similarly,
\begin{equation}
\begin{aligned}
P\left(U<\dfrac{f(Y_i)/P(A_i)}{M_i(\frac{1}
{B})}\right) &=\int_{A_i} \dfrac{f(y)/P(A_i)}{M_i(\frac{1}
{B})}\left( \frac{1}{B}\right) dy \\
\end{aligned}
\label{rejection sampling1i2}
\end{equation}

Using the results of Equation (\ref{rejection sampling1i1}), and Equation (\ref{rejection sampling1i2}) in Equation (\ref{rejection sampling1i}) we get

\begin{equation}
    \begin{aligned}
P(Y_i\leq x\,| Y_i  \hspace{.2cm}\text{accepted})
&=\dfrac{\displaystyle \int \textbf{I}_{(y\leq x)} P\left( U<\dfrac{f(y)/P(A_i)}{M_i(\frac{1}
{B})}\right) \left( \frac{1}{B}\right) dy}{\displaystyle 
 \int_{A_i} \dfrac{f(y)/P(A_i)}{M_i(\frac{1}
{B})}\left( \frac{1}{B}\right) dy.}\\
\end{aligned}
\label{rejection sampling1ic}  
\end{equation}

The cumulative distribution function of $X$ for the entire range is given by 
\begin{equation}
    P(X\leq x)= \int_{a}^{x} f(t)\,\,dt
    =  \left(\frac{1}{M} \int_{a}^{x} f(t)\,\,dt \right) \Bigg/ \left(\frac{1}{M}\right),
\label{rejection sampling2}  
\end{equation} 

where we can choose $M$ in such a way that

\[        M \geq \displaystyle \max_{x}\, \frac{f(x)}{p(x)} 
   =\displaystyle \max_{i} \biggl\{  \max_{x\in A_{i}} ~ \frac{f(x)}{p(x)}  \biggl\}
   = \displaystyle \max_{1\leq i \leq k}  \left[ H_{i} \Bigg/ \left( \frac{H_{i}}{B~\sum_{i=1}^{k}H_{i}} \right) \right]  = \displaystyle B\, \sum_{i=1}^{k}H_{i}. \]

Note that
\begingroup
\allowdisplaybreaks
\begin{align}
 \frac{1}{M} \int_{a}^{x} f(t)\,\,dt 
     &=  \sum_{i=1}^{k} \left[ \frac{1}{M} \int_{A_i} \textbf{I}_{(t\leq x)}\,\, \dfrac{f(t)}{p(t)} \,\, p(t) \,\, dt \right] \nonumber\\
     &=  \sum_{i=1}^{k} \left[ \frac{H_{i}}{M} \int_{A_i} \textbf{I}_{(t\leq x)}\,\, \dfrac{f(t)}{H_{i}\,p(t)} \,\, p(t) \,\, dt \right] \nonumber\\
     &=  \sum_{i=1}^{k} \left[ \frac{H_{i}}{M} \int_{A_i} \textbf{I}_{(t\leq x)}\,\, \dfrac{f(t)/P(A_{i})}{H_{i} /P(A_{i})} \,\,  \left(\frac{ \displaystyle \int_{A_{i}}p(t)\,dt}{p(t)}\right)\, \left(\frac{p(t)}{\displaystyle \int_{A_{i}}p(t)\,dt}\right) \,\, dt \right] \nonumber\\
     &=  \sum_{i=1}^{k} \left[ \frac{B\,H_{i}}{M} \int_{A_i} \textbf{I}_{(t\leq x)}\,\, \dfrac{f(t)/P(A_{i})}{B\,H_{i} /P(A_{i})} \,\,  \left(\frac{ 1}{\frac{1}{B}}\right)\, \left(\frac{1}{B}\right) \,\, dt \right] \nonumber\\
     &=  \sum_{i=1}^{k} \left[ \frac{B\,H_{i}}{M} \int_{A_i} \textbf{I}_{(t\leq x)}\,\, \dfrac{f(t)/P(A_{i})}{M_{i}\, \left(\frac{1}{B}\right)} \,\, \left(\frac{1}{B}\right) \,\, dt \right] \nonumber\\
     &=  \sum_{i=1}^{k} \left[ \frac{B\,H_{i}}{\displaystyle B\, \sum_{i=1}^{k}H_{i}} \int_{A_i} \textbf{I}_{(t\leq x)}\,\, \dfrac{f(t)/P(A_{i})}{M_{i}\, \left(\frac{1}{B}\right)} \,\, \left(\frac{1}{B}\right) \,\, dt \right] \nonumber\\
     &=  \sum_{i=1}^{k} \left[ \frac{H_{i}}{\displaystyle  \sum_{i=1}^{k}H_{i}} \frac{1}{M_{i}} \left( \frac{1}{1/M_{i}}  \right)
     \int_{A_i} \textbf{I}_{(t\leq x)}\,\, \dfrac{f(t)/P(A_{i})}{M_{i}\, \left(\frac{1}{B}\right)} \,\, \left(\frac{1}{B}\right) \,\, dt \right] \nonumber\\
     &=  \sum_{i=1}^{k} \left[ \frac{H_{i}}{\displaystyle  \sum_{i=1}^{k}H_{i}} \frac{P(A_{i})}{B\,H_{i}} \left( \frac{1}{1/M_{i}}  \right)
     \int_{A_i} \textbf{I}_{(t\leq x)}\,\, \dfrac{f(t)/P(A_{i})}{M_{i}\, \left(\frac{1}{B}\right)} \,\, \left(\frac{1}{B}\right) \,\, dt \right] \nonumber\\
     &=  \sum_{i=1}^{k} \left(\frac{P(A_{i})}{M} \right) \left[ 
     \dfrac{\displaystyle \int_{A_i} \textbf{I}_{(t\leq x)}\,\, P\left( U<\dfrac{f(t)/P(A_i)}{M_i(\frac{1}
{B})}\right) \,\, \left(\frac{1}{B}\right) \,\, dt }{\displaystyle \frac{1}{1/M_{i}} }
     \right] 
     \label{appendix rejection sampling21}
        \end{align}%
\endgroup

 Hence, using the Equation (\ref{rejection sampling1ic}) in Equation (\ref{rejection sampling2}) we get
\begin{equation}
      \dfrac{\displaystyle \frac{1}{M} \int_{a}^{x} f(t)\,\,dt }{\displaystyle \frac{1}{M}}  =  \sum_{i=1}^{k}  P(A_{i})  \left[ 
     \dfrac{\displaystyle \int_{A_i} \textbf{I}_{(t\leq x)}\,\, P\left( U<\dfrac{f(t)/P(A_i)}{M_i(\frac{1}
{B})}\right) \,\, \left(\frac{1}{B}\right) \,\, dt }{\displaystyle 
 \int_{A_i} \dfrac{f(t)/P(A_i)}{M_i(\frac{1}
{B})}\left( \frac{1}{B}\right) dt.}
     \right].  
\end{equation}


\end{document}